\documentclass[12pt]{article}
\usepackage{geometry}
\usepackage{lineno}

\usepackage[utf8]{inputenc} 
\usepackage{graphicx} 
\usepackage{amsmath} 
\usepackage{color,soul} 
\usepackage{rotating,tabularx} 
\usepackage{cite} 
\usepackage{comment}
\usepackage[T1]{fontenc}
\usepackage{authblk}
\usepackage{adjustbox}

\usepackage{setspace}
\usepackage{pdfpages}

\usepackage[hyphens]{url} 
\usepackage{cite}
\usepackage[hidelinks]{hyperref}
\hypersetup{breaklinks=true}

\title{Using protein blocks to build custom fragment libraries from protein structures}
\date{}
\author[1,2]{Surbhi Dhingra}
\author[1]{Stéphane Téletchéa}
\author[2]{Ramanathan Sowdhamini}
\author[1]{Yves-Henri Sanejouand}
\author[3,4]{Alexandre G. de Brevern}
\author[3,4,5]{Frédéric Cadet}
\author[1]{Bernard Offmann*}

\affil[1]{\small{Nantes Université, CNRS, US2B, UMR 6286, F-44000, Nantes, France}}
\affil[2]{Computational Approaches to Protein Science (CAPS), National Centre for Biological Sciences (NCBS), Tata Institute for Fundamental Research (TIFR), Bangalore 560-065, India}
\affil[3]{Université Paris Cité and Université de la Réunion, INSERM, EFS, BIGR U1134, DSIMB Bioinformatics team, F-75015 Paris, France}
\affil[4]{Université Paris Cité and Université de la Réunion, INSERM, EFS, BIGR U1134, DSIMB Bioinformatics team, F-97715 Saint Denis Messag, France}
\affil[5]{PEACCEL, AI for Biologics, F-75013 Paris, France}

\begin{document}
\pagestyle{empty}

\maketitle

\pagestyle{empty}
\section*{Abstract}

The remarkable structural diversity of modern proteins reflects millions of years of evolution, during which sequence space has expanded while many structural features remain conserved. This conservation is evident not only among homologous proteins but also in the recurrence of supersecondary motifs across unrelated proteins, underscoring the abundance and robustness of these structural units. Here, we present a novel pipeline for generating customized protein fragment libraries using protein blocks (PBs)—a structural alphabet that encodes local backbone conformations. Our method efficiently extracts structurally similar fragments from a curated, non-redundant protein structure database by converting three-dimensional structures into one-dimensional PB sequences. By integrating predicted PB sequences with the PB-ALIGN and PB-kPRED tools, our approach identifies relevant fragments independently of sequence homology. Fragment quality is further assessed using a new scoring function that combines secondary structure similarity and PB alignment metrics. The resulting libraries contain fragments of at least seven PBs (11 amino acid residues), covering over 70\% of the local backbone structure. Our results demonstrate that PBs enable efficient mining of high-quality structural fragments from diverse protein spaces, including proteins with disordered regions. The pipeline is accessible as an online tool (PB-Frag, \url{http://pbpred-us2b.univ-nantes.fr/pbfrag}).\\
\textbf{*Corresponding author:} \href{mailto:bernard.offmann@univ-nantes.fr}{bernard.offmann@univ-nantes.fr}\\
\textbf{Keywords:} Fragment-based design, protein structure prediction, structural alphabet, local backbone conformations, protein blocks\\
\textbf{Supplementary information:} Electronic Supplementary Information are provided.

\pagestyle{empty}
\newpage
\pagestyle{plain}

\section{Introduction}

Fragment-based design (FBD) is a foundational methodology in protein structure prediction and design. It involves constructing complete protein models by assembling short, local structural fragments derived from known protein structures \cite{Rohl2004, Zhang2005, Xu2012, dhingra2020glance}. This approach facilitates efficient exploration of conformational space while incorporating both evolutionary and geometric constraints.

FBD approaches primarily depend on mining structural space through local sequence comparisons. The underlying principle is that local protein sequence patterns tend to exhibit characteristic structural features \cite{Simons1997}. This observation led to the hypothesis that the local conformations of a given protein sequence can be reliably inferred by identifying fragments that have been structurally characterised and that have similar local sequence motifs in existing protein structure databases. \cite{Simons1997,Kandathil2018}. Typically, FBD approaches identify multiple fragments that cover each position on the target protein sequence, which are then filtered to select the most representative candidates based on various scoring criteria. The length of these fragments varies depending on the algorithm, but commonly falls within a range of up to 20 residues \cite{DeOliveira2015}. However, accurate models have also been achieved using fragments as short as three residues \cite{Bonneau2001,Gront2011}.

Fragment-based approaches are particularly advantageous because they restrict the dimensionality of the conformational search space by limiting the number of fragments considered at each sequence position. This restriction also presents a significant limitation: these algorithms may fail to adequately explore alternative conformations for a given  sequence \cite{Kandathil2016}. To address this drawback, recent efforts have focused on redesigning fragment search heuristics to enhance conformational diversity\cite{Kandathil2018}.

At the same time, advances in generative protein modelling, such as ESMFold \cite{lindert2012emfold-25f} and OmegaFold \cite{Wu2022:OmegaFold}, have demonstrated the power of data-driven approaches in capturing local and global structural features, challenging the traditional reliance on sequential heuristics.
Additionally, attention-based architectures like ProteinMPNN \cite{dauparas2022robust-163}, and language models, such as ProGen2 \cite{nijkamp2023progen2-754}, have shown promising capabilities in protein sequence design. These developments underscore the increasing synergy between machine learning and fragment-based methods.
Together, these innovations point to the exciting potential of integrating structural alphabets into deep learning pipelines, paving the way for more efficient, interpretable and generalisable protein modelling frameworks. Nevertheless, further supervised analyses are needed to enhance the explainability of these models.

Two main types of fragment search approaches have been used. The first is the classical sequence-based search, which uses local sequence similarity search algorithms to identify structural fragments from known protein structures. The second is a structure-based search, which relies on local structural similarity search algorithms to find such fragments (for a review see \cite{dhingra2020glance}). Only a few instances of structure-based fragment generation have been documented in recent years. One notable example is SA-Frag \cite{Shen2013}, which uses a type of structural alphabet (SA) to construct fragment libraries. This protocol compares local profiles between target and template structures based on predicted SA sequences. The study successfully introduced the concept of SAs into protein structure prediction, although it has not yet achieved the performance of sequence-based methods \cite{Abbass2015}. This gap highlights the need for further exploration of the usage of SAs in the field of structure prediction.

A typical structural alphabet consists of a limited set of short structural prototypes, derived by clustering recurrent structural motifs found in existing protein structures. These prototypes provide an effective mean to approximate the local backbone conformations of proteins \cite{Unger1989,Camproux2004,Li2008, Brevern2000,Brevern2005,Valadie2010}. One well-established example is the protein blocks (PBs) structural alphabet, which comprises 16 distinct structural prototypes labeled from \emph{a} to \emph{p}. Each PB represents a segment of 5 residues that is recurrently observed in local protein structures \cite{Brevern2000,Brevern2005}. These prototypes were identified by analysing and clustering patterns of dihedral angles ($\phi$ and $\psi$) spanning over five consecutive residues. For a comprehensive review about protein blocks see \cite{offmann2007review, Valadie2010}.

\begin{figure}[!h]
\centerline{\includegraphics[width=0.9\textwidth]{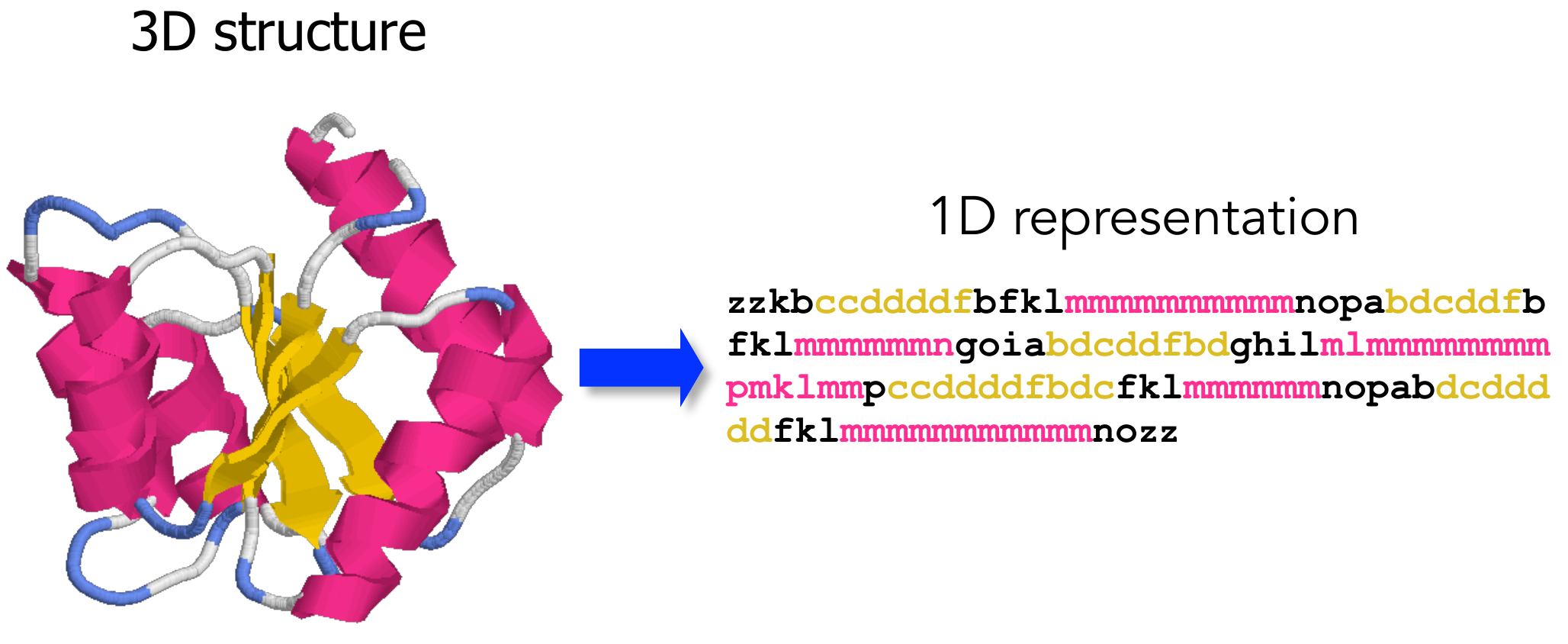}}
\caption{\textbf{Principle of precise encoding of a protein 3D structure information into a simplified 1D representation using protein blocks}. Each PB is a structural motif that spans over 5 residues and is represented by a letter between \emph{a} to \emph{p}. For example, PB \emph{m} (here letter $m$) is a structural motif that is typical of $\alpha$ helical regions. Likewise, PB \emph{d} is typical of the central part of a $\beta$ strand. PB sequence on the right is coloured according the regular secondary structures in the 3D structure.}\label{fig:01}
\end{figure}

By applying PBs, the three-dimensional atomic coordinates of a protein structure can be converted into a one-dimensional PB sequence through a process known as PB assignment (see Figure~\ref{fig:01}). This resulting 1D PB sequence serves as a compressed yet accurate representation of the local protein structure.

PB sequences facilitate protein structure comparison by enabling approaches similar to sequence alignment. To this end, a PB sequence alignment methodology, PB-ALIGN, was developed. PB-ALIGN utilises a PB substitution matrix and a dynamic programming algorithm to align PB sequences \cite{tyagi2006substitution}. It is available as a web server, the Protein Block Expert (PBE) -- \url{https://pbpred-us2b.univ-nantes.fr/pbe/?page_id=12} -- and supports both local and global structure alignment \cite{Tyagi2006PBE}.

Another application based on PBs is PB-kPRED, which predicts local backbone conformation from a protein sequence using a knowledge-based scoring function without relying on secondary structure and sequence alignment profiles\cite{Vetrivel2017}. This algorithm is accessible as a web-based tool (PB-kPRED, \href{https://pbpred-us2b.univ-nantes.fr/kpred/}{https://pbpred-us2b.univ-nantes.fr/kpred/}).

In this study, we utilised PB-ALIGN and PB-kPRED tools to systematically mine recurrent protein structural motifs and construct query-based fragment libraries. These libraries were validated through sequence and secondary structural comparisons. Additionally, we developped a new scoring function to identify high quality fragment, specially those with backbone conformations most closely matching the target sequence.

Our results underline the interest of this PB-based approach to efficiently extract large numbers of high-quality structural fragments from a database of unrelated protein structures. These customized fragment libraries offer new opportunities for fragment assembly methods in protein design.

\section{Materials and Methods}

\subsection{Curated Template Database}

A non-redundant database of protein structures was set up by downloading protein chain entries from RCSB Protein Data Bank (www.rcsb.org) \cite{Berman2000}. The following selection criteria were applied: (a) experimental method, X-ray crystallography, (b) resolution $\leq$ 3\AA, (c) R-factor $\leq$ 0.2, and (d) protein chain length $\geq$ 40 residues. This yielded 23,989 unique protein chains. The chains were clustered at 30\% sequence identity using the KClust algorithm \cite{Hauser2013}, resulting in a total of 7,632 clusters. Proteins with chain breaks were excluded, consolidating the database to 5,391 unique chains, designated hereupon as the PDB30 database. PB sequences were assigned to each chain using an in-house script, and secondary structure assignments were generated with Pdb-tools \cite{Rodrigues2018}.

\subsection{Query Dataset}

The query dataset was adapted from a previous study focused on fragment library generation \cite{DeOliveira2015}. It comprises 43 query protein structures ranging from 59 to 508 residues in length (see Table \ref{Tab:01}). The dataset was designed to represent four major SCOP classes, i.e., all-$\alpha$, all-$\beta$, $\alpha$/$\beta$ and $\alpha$+$\beta$. Each protein in the dataset is a homomer and monomeric units were used for the analysis.

\subsection{Protein Block Prediction}

The knowledge-based tool PB-kPRED \cite{Vetrivel2017} was used to predict PB sequences for each query protein. To avoid bias, all templates from PB-kPRED internal database sharing $\geq$30\% sequence identity with the query protein were removed. Table \ref{Tab:01}\vspace*{1pt} presents the dataset along with PB prediction accuracy. Secondary structure predictions were performed using PSIPRED \cite{Jones1999, mcguffin2000psipred}.

\begin{table}[!h]
\caption{{\bf The query dataset and its characteristics.}} 
\label{Tab:01}
\small
The table gives for each entry, its SCOP class, its length in number of amino acid residues (AA) and observed in its experimental structure (PDB), and its accuracy in terms of PB-kPRED's PB prediction (\%).\\
\begin{center}
{\begin{tabular}{@{}llllll@{}}
\hline 
PDB id & SCOP Class & Length (AA) & Length (PDB) & Accuracy (\%)\\
\hline
1AIL & all-$\alpha$ & 73 & 70 & 49.8\\
1RRO & all-$\alpha$ & 108 & 108 & 27.6\\
1U61 & all-$\alpha$ & 138 & 127 & 24.3\\
1SL8 & all-$\alpha$ & 191 & 181 & 35.8\\
1QUU & all-$\alpha$ & 250 & 248 & 68.1\\
1T5J & all-$\alpha$ & 313 & 301 & 30.7\\
1PO5 & all-$\alpha$ & 476 & 465 & 29.2\\
1MHN & all-$\beta$ & 59 & 59 & 69.0\\
1TEN & all-$\beta$ & 90 & 90 & 35.3\\
2G1L & all-$\beta$ & 104 & 103 & 44.1\\
1IFR & all-$\beta$ & 121 & 113 & 49.8\\
1BFG & all-$\beta$ & 146 & 126 & 45.3\\
2FR2 & all-$\beta$ & 172 & 161 & 32.6\\
1EE6 & all-$\beta$ & 197 & 197 & 49.7\\
1UAI & all-$\beta$ & 224 & 223 & 27.1\\
2C9A & all-$\beta$ & 259 & 259 & 74.3\\
1O4Y & all-$\beta$ & 288 & 270 & 31.2\\
1HG8 & all-$\beta$ & 349 & 349 & 40.7\\
1NKG & all-$\beta$ & 508 & 508 & 74.2\\
1VJW & $\alpha$+$\beta$ & 60 & 59 & 33.5\\
1MWP & $\alpha$+$\beta$ & 96 & 96 & 75.8\\
1GNU & $\alpha$+$\beta$ & 117 & 117 & 21.4\\
1R9H & $\alpha$+$\beta$ & 135 & 118 & 28.9\\
206L & $\alpha$+$\beta$ & 164 & 162 & 79.1\\
2FS3 & $\alpha$+$\beta$ & 282 & 280 & No Pred\\
1DZF & $\alpha$+$\beta$ & 215 & 211 & 39.1\\
1DXJ & $\alpha$+$\beta$ & 242 & 242 & 49.2\\
1MAT & $\alpha$+$\beta$ & 264 & 263 & 30.6\\
1JKS & $\alpha$+$\beta$ & 294 & 280 & 33.0\\
1MC4 & $\alpha$+$\beta$ & 370 & 369 & 29.0\\
2FKF & $\alpha$+$\beta$ & 462 & 455 & 32.5\\
1H75 & $\alpha$/$\beta$ & 81 & 76 & 42.7\\
1IU9 & $\alpha$/$\beta$ & 111 & 111 & 71.5\\
1E6K & $\alpha$/$\beta$ & 130 & 130 & 52.0\\
1P90 & $\alpha$/$\beta$ & 145 & 123 & 45.6\\
1FTG & $\alpha$/$\beta$ & 168 & 168 & 40.1\\
1QCY & $\alpha$/$\beta$ & 193 & 193 & 55.1\\
2A14 & $\alpha$/$\beta$ & 263 & 257 & 34.9\\
1IZZ & $\alpha$/$\beta$ & 283 & 276 & 31.6\\
1QUE & $\alpha$/$\beta$ & 303 & 303 & 45.2\\
1KRM & $\alpha$/$\beta$ & 356 & 349 & 34.7\\
3BSG & $\alpha$/$\beta$ & 414 & 404 & 76.7\\
1PGN & $\alpha$/$\beta$ & 482 & 473 & 63.0\\
\hline
\end{tabular}}
\end{center}
\end{table}

\begin{figure*}[!ht]
\centerline{\includegraphics[width=130mm]{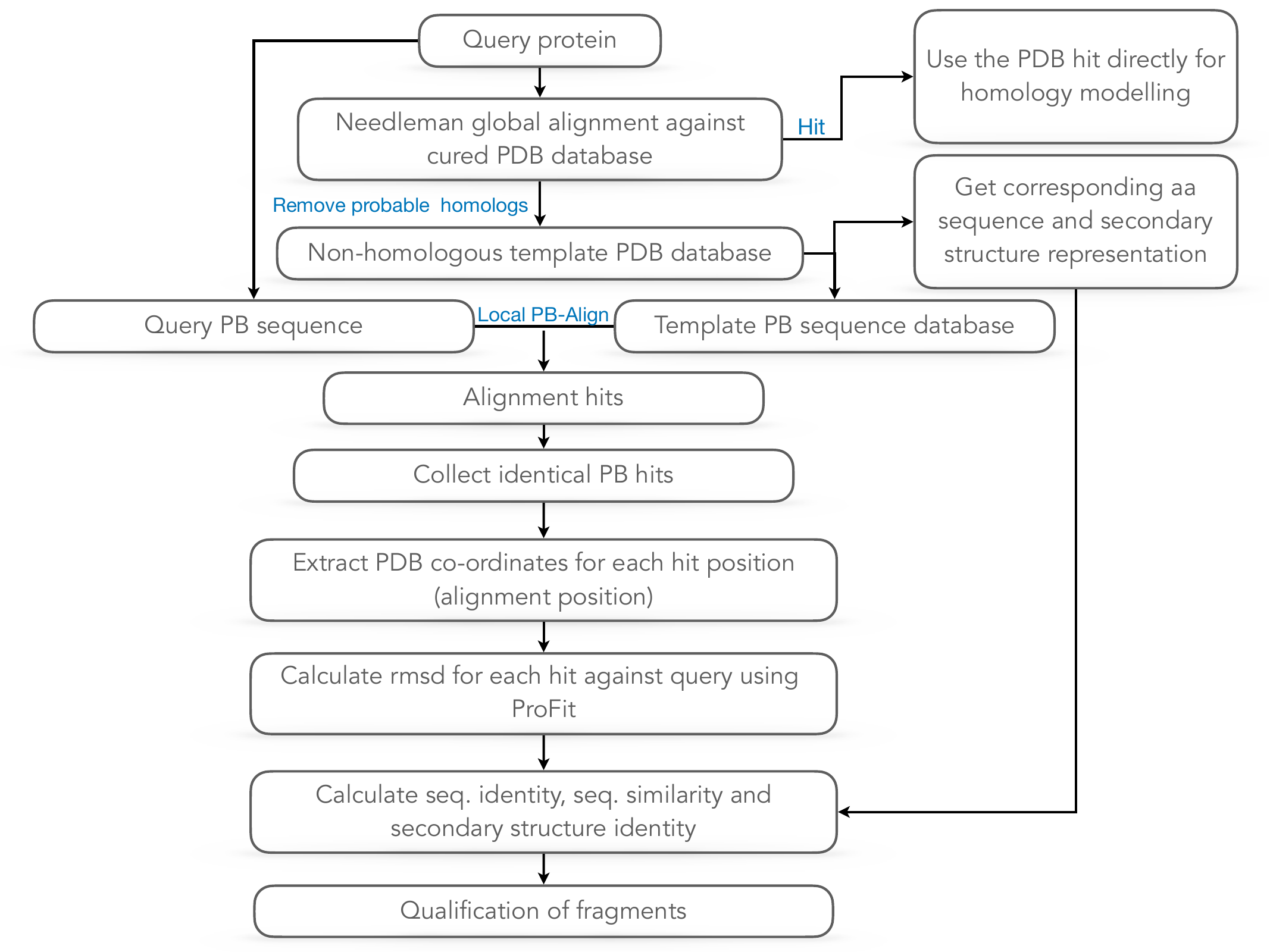}}
\caption{\textbf{PB-based fragment generation and evaluation pipeline.} Schematic overview of the protein blocks-based fragment generation and evaluation workflow. The process begins with a query protein, for which homologous sequences are excluded from the curated structural database (PDB30). Only non-redundant entries with less than 30\% sequence identity are retained. The query is then converted into a PB sequence using PB-kPRED prediction tool \cite{Vetrivel2017}, which predicts local backbone conformations in a simplified one-dimensional (1D) representation. This PB sequence is then locally aligned to template PB sequences in the database using the PB-ALIGN \cite{tyagi2006substitution} to identify structurally similar regions. For each alignment, the corresponding 3D coordinates are extracted to generate candidate fragments of at least 11 residues. These fragments are evaluated by structural superposition with the query using the root mean square deviation (\textit{RMSD}) method, as well as by comparing sequence identity, similarity and secondary structure identity. A custom scoring function ($atan\ score$, see equation \ref{equation:atan}) integrates these metrics to assess fragment quality. Fragments exceeding the threshold are retained, resulting in a targeted library of structural building blocks for downstream modelling applications. }\label{fig:02}
\end{figure*}

\subsection{Fragment Mining}

Fragments were extracted from the PDB30 database. Any template sequence sharing $\geq$30\% sequence identity with a query was identified by global alignment \cite{Needlemanlnd1970} and excluded prior to analysis. Local PB alignments were performed using the PB-ALIGN tool \cite{tyagi2006substitution}, enabling 1D structural comparisons and identification of local conformations. The minimum fragment length was set to 7 PBs (11 residues). The overall fragment generation process is summarised in Figure~\ref{fig:02}, which outline the pipeline used in this study.

\subsection{Fragment Quality Assessment}

Fragments and query structures were superimposed at each position using the Bio module in biopython to calculate \textit{RMSD} as a quality criterion. Coverage was defined as the number of positions in the query sequence for which at least one fragment was identified by the  pipeline. Additional assessments included sequence identity, sequence similarity, and secondary structure identity between fragment hits and the target sequence. A scoring function, termed the $atan\ score$, was developed based on observed sequence variance. The $atan\ score$ integrates secondary structural identity ($ssID$) calculations and the normalised PB-ALIGN score ($nscore$) as follows: 
\begin{equation}
    atan\ score=atan\left(nscore \cdot \left(\frac{ssID}{100}\right)^2\right)
    \label{equation:atan}
\end{equation}

\begin{figure}[h]
\centerline{\includegraphics[width=0.65\textwidth]{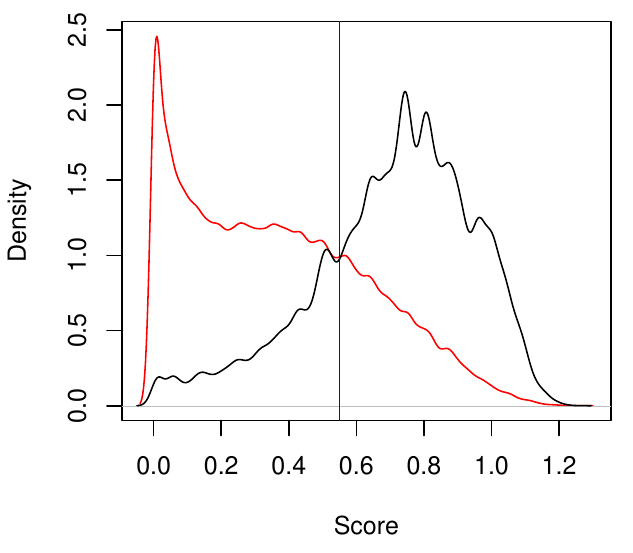}}
\caption{{\bf Distribution of fragment atan scores.} Probability density plots of the atan score (see Methods) for all fragments. Scores for fragments with \textit{RMSD} $\leq$2.5\AA\ are shown in black, while those with \textit{RMSD}$>$2.5\AA\ are shown in red. The blue line indicates the cutoff value of 0.55, which best separates high- and low-quality fragments. This threshold was subsequently used to calculate sensitivity and specificity.}\label{fig:03}
\end{figure}

Large-scale analysis of fragment data and the distribution of atan score revealed that fragments with $atan\ score \geq0.55$ exhibit high backbone similarity to the query structure (Figure \ref{fig:03}).

\section{Results}

\subsection{Template database}

The final template database (PDB30) comprised 5,391 protein chains, each with less than 30\% sequence identity and a resolution better than 3\AA{}. The distribution of secondary structural elements showed that it is composed of 44.8\% $\alpha$-helices, 27.8\% of $\beta$-sheets and 27.4\% coils. In terms of PBs, the distribution was 29.9\% PB \textit{m} (representing the central part of $\alpha$-helices), 19.1\% PB \textit{d} (the central part of a $\beta$-strand) and 51.0\%  other PBs (mainly coils). This closely reflects the typical distribution of regular and irregular secondary structures observed in proteins \cite{Valadie2010}. 

The PDB30 database includes representatives from 10 out of 12 SCOP classes, with the majority belonging to the four main SCOP classes. Out of the 5,391 protein chains, 1,962 could not be assigned a corresponding SCOP class, likely due to delays in the synchronisation of structural annotation data across databases. 

\subsection{Fragment mining and generation}

To minimise bias from close homologs, any template sequence sharing >30\% sequence identity a query was dynamically removed from the PDB30 databank during the analysis (see Figure \ref{fig:02}). As a result,  over 99\% of the remaining template sequences shared less than 20\% sequence identity with the queries. The median sequence identity was 12.4\%, closely matching the theoretical value of 12\% expected for random sequence alignment \cite{rost1999twilight-117}.

On average, the pipeline generated approximately $\sim$53k hits per query protein, with a minimum of 32,921 hits for Receptor-type Tyrosine-protein phosphatase $\mu$ (PDB ID: 2C9A) and maximum of 66,034 hits for Rpb5 protein (PDB ID: 1DZF). Detailed counts of total hits and the number of hits with $atan\ score \geq0.55$ are provided in Supplementary Table 1.

The majority of fragments were 15 $\pm$ 5 in length across all query proteins, although fragments up to 100 residues were occasionally observed (see Supplementary Table 2). Figure~\ref{fig:04} depicts a barplot of the overall sequence coverage for each query protein: blue bars indicate coverage using the complete set of fragments, while green bars represent coverage after filtering for fragments with an $atan\ score \geq0.55$.

\begin{figure*}[h]
\centerline{\includegraphics[width=1.0\textwidth]{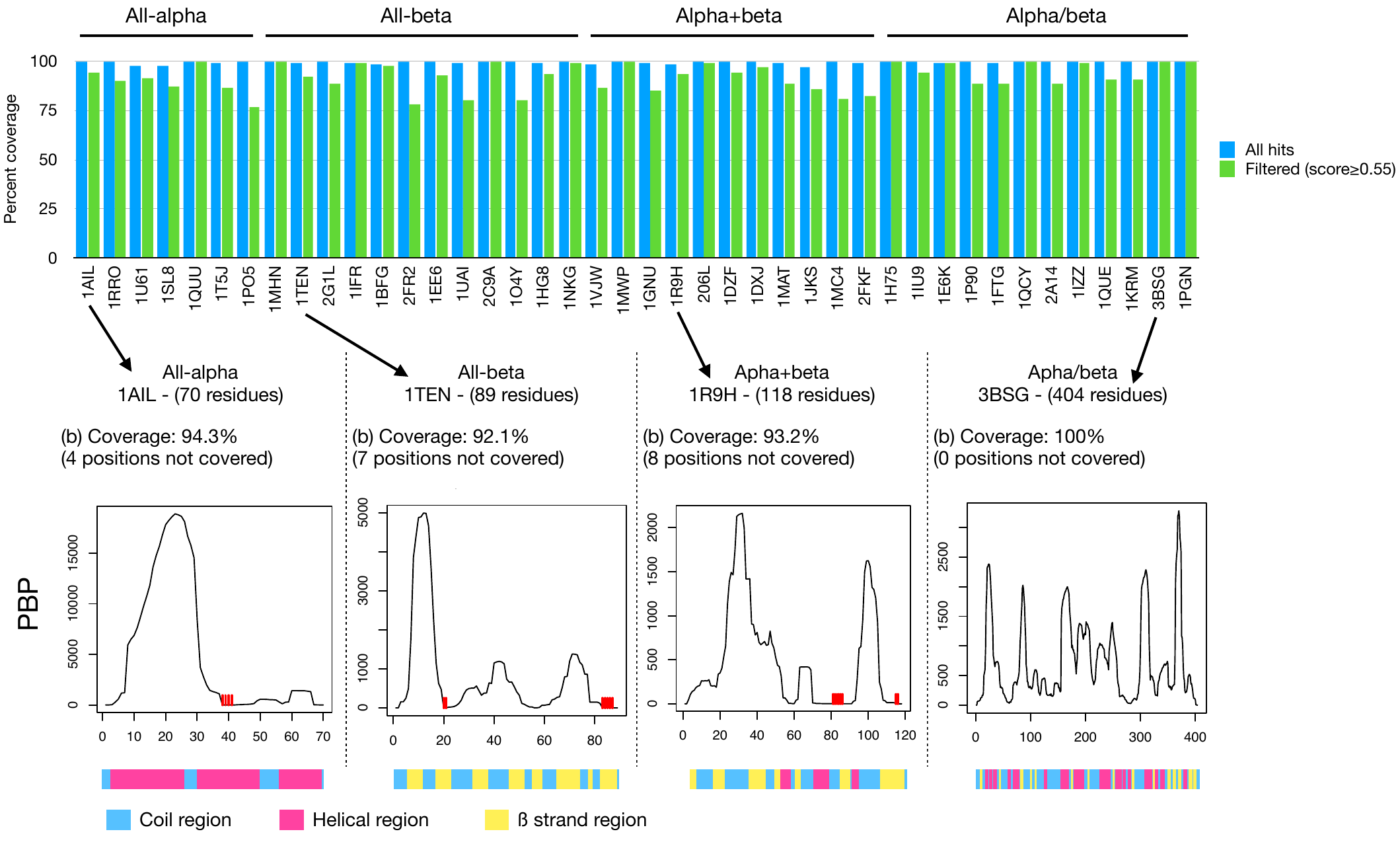}}
\caption{{\bf Coverage analysis of generated fragments.} {\it Top}: Bar plot showing the percentage of sequence positions covered by at least one fragment for each query protein (x-axis are the labels for the PDB codes of the queries). Results for all fragments and for fragments with $atan\ score\geq 0.55$ are shown in blue and green respectively. Query proteins are grouped by SCOP class. {\it Bottom}: Detailed coverage profiles for 4 query examples. The x-axis indicates residue positions and the y-axis shows the number of fragments covering each position. Residues not covered by any fragment highlighted in red.}\label{fig:04}
\end{figure*}

Some regions of the protein are more densely populated with fragments than the others, as illustrated in Figure~\ref{fig:04} for representative queries from each SCOP class. This pattern reflects a higher natural abundance of specific structural motifs. The distribution of fragment coverage along the length of each query protein highlights these differences, with the most highly covered regions typically corresponding to canonical secondary structures elements. 

A web server -- PB-Frag -- which implements the methodology, is available (\url{http://pbpred-us2b.univ-nantes.fr/pbfrag}). It identifies and extracts structurally similar fragments for any input query protein sequence and corresponding predicted PB sequence. The user is required to run PB-kPRED (\url{https://pbpred-us2b.univ-nantes.fr/kpred/}) prior to the submitting to PB-Frag in order to get a predicted PB sequence. Additionally, we are providing a complementary tool, PB-Extractor (\url{https://pbpred-us2b.univ-nantes.fr/pbe/?page_id=206}), that helps users in mining the PDB to retrieve atomic coordinates of fragments matching a given PB sequence.

\subsection{Assessment of fragment quality}

\begin{figure*}[ht]
\centerline{\includegraphics[width=1.0\textwidth]{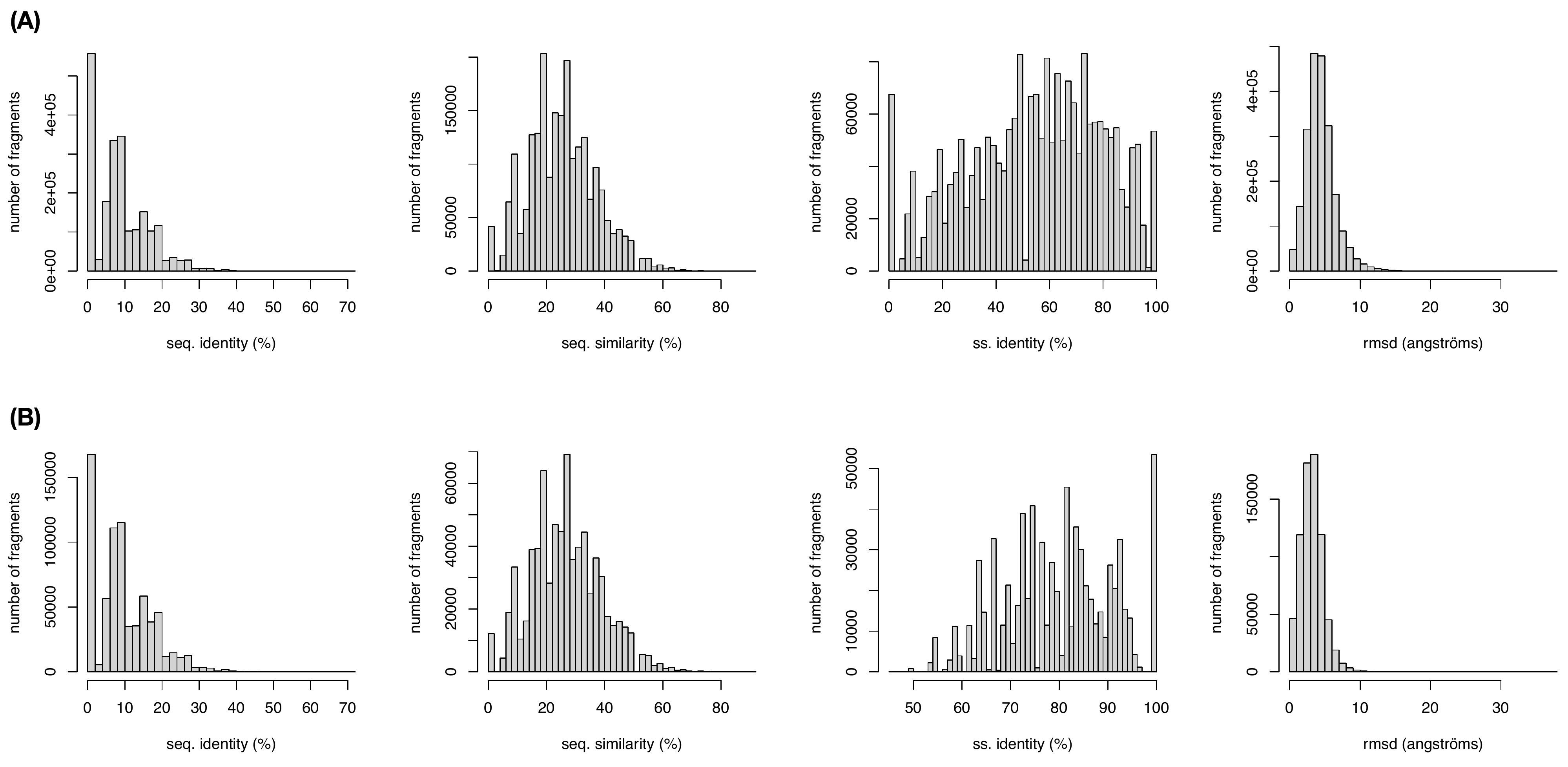}}
\caption{{\bf Qualitative analysis of fragment generated by the pipeline.} The histograms display the distribution of sequence identity, sequence similarity, secondary structure identity and \textit{RMSD}. (A) Distribution for all fragments generated by the pipeline. (B) Distribution for the fragments with $atan\ score>0.55$.}\label{fig:05}
\end{figure*}

For each fragment hit, \textit{RMSD} was calculated relative to the corresponding position in the query protein. Amino acid sequence identity, sequence similarity, and secondary structure identity ($ssID$) were also determined for all fragments at their respective query positions. Figure~\ref{fig:05} depicts the overall distributions of these metrics for all the fragment hits (Figure \ref{fig:05}A) and for those exceeding the $atan\ score$ cutoff (Figure \ref{fig:05}B). 

Notably, the distribution of amino acid sequence identity is dominated by fragments with no sequence identity (0\%) to the query, while sequence similarity is slightly higher but still skewed towards lower values. This indicates that, even at the local level, most fragments are not closely related to the query in terms of amino acid sequence.. In contrast, the distribution of secondary structure identity, particularly for fragments above the $atan\ score$ threshold, show a marked increase in matches (see Figure \ref{fig:05}B). This difference is expected, as secondary structure is classified into three states, compared to the 20 possible amino acids. 

Many fragments shared identical secondary structural features with the query protein, reflecting the design of PBs to provide one dimensional description of the local protein backbone. This property makes secondary structure identity an effective and objective criterion for assessing and qualifying fragment quality. 

\begin{figure}[h]
\centerline{\includegraphics[width=1.0\textwidth]{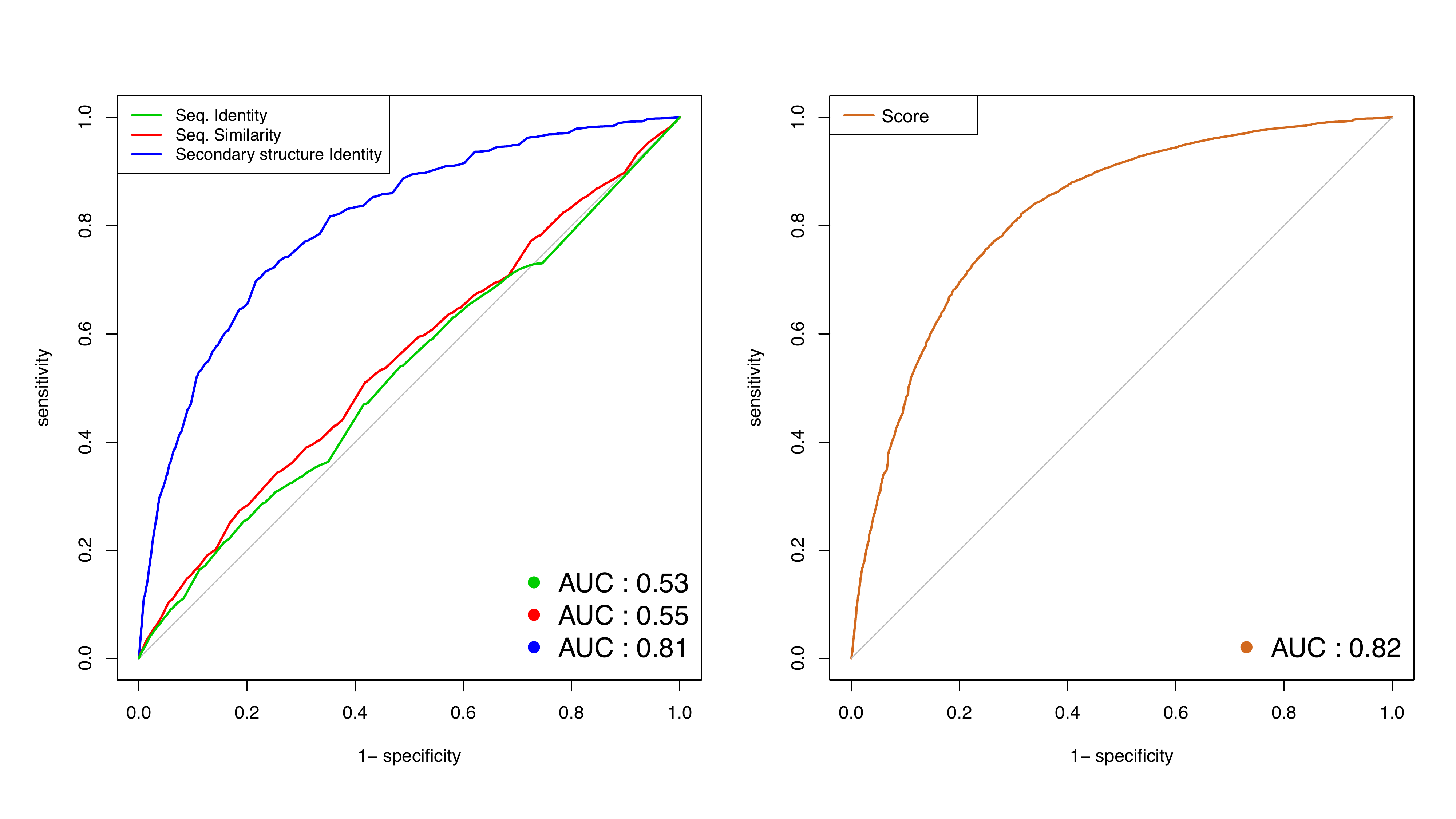}}
\caption{\textbf{ROC analysis of four objective criteria for fragment selection}. ROC curves are shown for all fragments with \textit{RMSD}below the 2.5\AA{} cut-off, evaluating four criteria: (i) amino acid sequence identity (green), (ii) amino acid sequence similarity (red), (iii) secondary structure identity (blue), and (iv) the $atan\ score$ (brown). Sequence identity and similarity display AUC values around 50\%, indicating little correlation with fragment quality. In contrast, secondary structure identity and the $atan\ score$ both achieve AUC values above 80\%, highlighting their effectiveness in identifying structurally relevant fragments.}\label{fig:06}
\end{figure}

An ROC curve analysis (see Figure~\ref{fig:06}) using an \textit{RMSD} threshold of (2.5\AA{}) confirmed that secondary structure identity is the most effective criterion among those tested for prioritising fragments. A similar trend was observed for the $atan\ score$ in relation to \textit{RMSD}. Sensitivity and specificity curves for individual SCOP classes are provided in Supplementary Figures 2a-d).

\begin{figure}[h]
\includegraphics[width=1.0\textwidth]{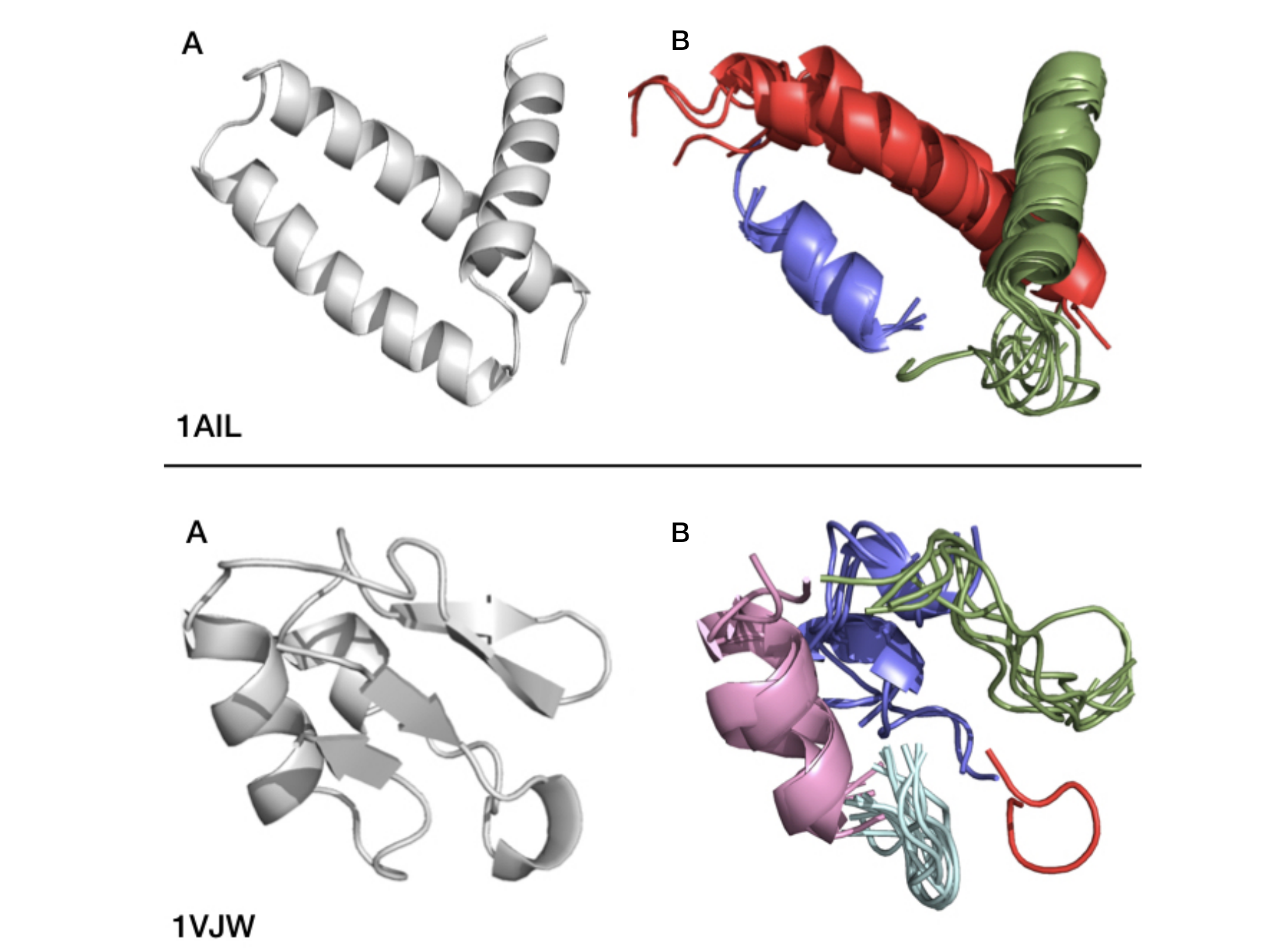}
\caption{{\bf Superimposition of generated fragments  onto original query structures.} Shown are (A) the original structures of two test proteins -  nonstructural protein 1 from influenza A virus (PDB ID: 1AIL) and the 1[4Fe-4S] ferredoxin from \textit{Thermotoga maritima} (PDB ID: 1VJW) - together with (B) the superimposed fragments retained for each.}\label{fig:07}
\end{figure}

Visual inspection of the fragments using PyMOL \cite{delano2002pymol} demonstrates that the PB-based fragment generation pipeline effectively preserves a large portion of local protein structural features. An example is shown in Figure~\ref{fig:07}, where panels A, B and C display the original structure, superimposed fragments generated by Protein Block Assignment (PBA), and those generated by Protein Block Prediction (PBP), respectively. These clearly illustrate that the PB-based approach can reliably extract structurally similar local regions for a given protein sequence.

The quality of fragments in regions with regular secondary structures ($\alpha$-helices and $\beta$-sheets) was compared to unstructured (coil) regions (see Supplementary Table 3). Overall, fragments generated for $\alpha$-helical regions exhibited lower \textit{RMSD} values than those for non-helical regions, indicating higher structural accuracy. In contrasts,  fragments generated for $\beta$-strands and coil regions showed similar \textit{RMSD} distributions (see Supplementary Table 3). Detailed distributions of \textit{RMSD} per query protein are featured in Supplementary Figures 3a-d. These illustrate that good quality fragments are obtained for almost all positions of a query protein.

\section{Discussion}

This study demonstrates the use of Protein Blocks as an efficient tool for extracting structurally similar fragments from protein strutures, even in the absence of sequence homologs. Our pipeline leverages PB-based alignments to detect local structural motifs directly from a curated non-redundant structural space (PBD30), shifting the focus from sequence-based to structure-based fragment selection. by facilitating the mapping of redundant structural motifs in protein space. The approach efficiently recovers a large pool of structural structural stretches of varying lengths for each query, notably including loop regions that connect regular secondary structures, regions often treated separately in template-free modelling protocols \cite{Maurice2014, Khor2015}.

PB sequence for all query proteins were predicted using PB-kPRED, with the exception of one case (Major capsid protein from bacteriophage, PDB ID: 2FS3). The average PB-prediction accuracy was 56.7\% for 42 out of 43 targets, sufficient to guide effective fragment selection. PB-kPRED is trained on pentameric units, and the high probability of pentamer occurence in natural sequences minimizes the impact of excluding full-length homologs from the PB-PentaDB during prediction. Ideally, the PB-kPRED tool would be a non-modified algorithm. However, to avoid distorting the query, the counterparts of each query were removed from the PB-PentaDB database before the PB sequence was predicted using PB-kPRED. This reduced the accuracy of the PB sequence predictions, which would not be the case with conventional use of the tool.

To further prevent bias, any potential homologs were removed from the PDB30 database before fragment generation, ensuring objectivity in fragment quality assessment 

Our analysis shows that structurally similar fragments can be identified even when sequence identity and similarity are low. In contrast, secondary structure identity between fragments and queries remains relevant, and higher secondary structure identity correlates with lower \textit{RMSD} values. This is expected, as PBs provide a detailed, 16-state representation of the protein backbone, offering finer resolution than traditional three-state secondary structure descriptions, especially for coil regions. In some cases, PB prediction accuracy (\textit{Q}${16}$) reached up to 70\%, surpassing typical three-state (\textit{Q}${3}$) prediction rates and highlighting the advantage of PB-based searches.

Fragment quality was primarily assessed by \textit{RMSD}, with additional validation using a composite $atan\ score$ that combines secondary structure identity and normalized PB-Align scores. While our pipeline’s efficiency is comparable to existing methods such as HHFrag and NNMake that report efficiency of 62.16\% and 38.17\% respectively, direct comparisons are challenging due to differences in scoring criteria and fragment selection strategies \cite{DeOliveira2015}. Notably, higher performance reported by methods like HHFrag can be attributed to the inclusion of sequence homologs in their databases \cite{DeOliveira2015}, a factor we explicitly controlled for in our protocol.


SAFrag, another structural alphabet–based fragment mining tool, reported 86.7\% high-quality fragments\cite{Shen2013}. Similar to HHFrag, it employs HMM-based profile-profile comparisons to identify fragment hits. Notably, SAFrag uses two structural databases -- PDB25 and PDB50 -- for fragment generation. Its higher coverage largely results from including the target structure and structural homologs in its template database. 

Our methodology employs relatively simple scoring functions for fragment generation, reflecting the flexible nature of PB-based fragments, which are generated in overlapping segments of varying lengths rather than as a fixed number per position. In contrast, methods like HHFrag and NNMake define a set number of fragments per position (averaging 10 and 200, respectively), enabling more uniform coverage. Consequently, our approach results in an uneven distribution of fragments across secondary structure regions and sequence positions (see Supplementary Tables 3 and 4; Figure~\ref{fig:04}), complicating fragment quality assessment. Importantly, a higher number of hits does not necessarily correspond to higher fragment quality, as observed for all-$\beta$ class queries. Consistent with previous studies and the known distribution of dihedral angles, fragments from helical regions consistently exhibited better quality (lower \textit{RMSD}) compared to those from non-helical regions.

Reduced PB prediction accuracy affected overall coverage, but this is a limitation of the modified PB-PentaDB used in this study. In practical applications, restoring the full PB-PentaDB is expected to improve both coverage and fragment quality.

The results are promising and suggest several avenues for refinement. Expanding the PDB30 dataset and updating the PB-PentaDB could enhance fragment diversity and PB-kPRED accuracy. Additionally, reducing the minimum fragment length may further improve precision, although literature supports 10–11 residue fragments as optimal \cite{bornot2011, Xu2012, Xu2012b}.

Our methodology diverges from conventional fragment mining by using PB sequences instead of amino acid sequences, challenging the reliance on sequence similarity for identifying structural features. Given the limited number of protein fold patterns compared to the vast sequence space, focusing on structural motifs significantly broadens the search landscape. Unlike SA-Frag, our pipeline constructs pairwise PB sequence alignments without length constraints or reliance on homologous sequences.

The pipeline is available as a web server, PB-Frag (\url{http://pbpred-us2b.univ-nantes.fr/pbfrag}), which identifies and extracts structurally similar fragments for any protein sequence. The server provides interactive plots, including coverage and secondary structure identity, and allows users to download customized fragment libraries with quality indicators. In addition, a complementary tool, PB-Extractor, assists users in mining the PDB to retrieve atomic coordinates of fragments matching a given PB sequence (\url{https://pbpred-us2b.univ-nantes.fr/pbe/?page_id=206}).  These resources are well suited for applications in protein engineering, chimeragenesis, and \textit{de novo} protein structure prediction.

\section{Conclusion}

Our results demonstrate that structural alphabets, such as Protein Blocks, are powerful tools for mapping and recovering structurally redundant regions from representative protein databases. Compared to amino acid sequence-based fragment libraries, SAs enable access to a broader conformational space, often overlooked in sequence-based approaches. This expands the search space while maintaining the ability to capture the native fold of target proteins. PB-mined fragments can also reveal subtle backbone variations that have evolved to enhance protein stability or function across different folds. Importantly, our approach enables the identification of structural homologs among proteins with low or no sequence similarity, and readily generates fragments covering the full length of small proteins, thereby facilitating structure prediction protocols.

Because PBs represent local conformations as one-dimensional sequences, they increase the likelihood of retrieving fragments with similar folds compared to amino acid sequence alignments. The protocol also allows extraction of longer fragments, potentially encompassing entire domains. Overall, PBs offer a promising foundation for protein structure prediction, using local conformations as a starting point.

Recent advances in FBD have been markedly accelerated by the integration of deep learning techniques, particularly diffusion models and autoregressive frameworks. Notable examples include RFdiffusion, a diffusion-based model for de novo protein backbone generation \cite{watson2023novo}, PepHAR, a hotspot-guided peptide design method leveraging multi-fragment autoregressive extension \cite{li2024hotspot} and FrameFlow, a fast protein backbone generation framework based on SE(3) flow matching \cite{yim2024improved}. These data-driven approaches enhance the precision, diversity, and scalability of protein design.

Importantly, our PB-based strategy can be seamlessly incorporated into classical FBD pipelines. Following the definition of a target architecture—such as a Rossmann-like $\alpha$/$\beta$ domain or a helical bundle—deep learning-based models can be employed during the fragment selection and backbone assembly stage, augmenting traditional methodologies (e.g., \cite{huang2022backbone}). Subsequent steps include sequence optimization, where side-chain packing and energetics are refined, and in silico validation, during which the designed structure is assessed using conformational sampling and state-of-the-art structure prediction tools such as AlphaFold2 \cite{jumper2021highly} to ensure folding competence and structural integrity. Ultimately, experimental validation—through expression, folding assays, and high-resolution structural determination—remains essential to confirm design success and guide further improvements.

\section*{Acknowledgements}

The authors thank Prof Narayanaswamy Srinivasan for fruitful discussions on this work.
These works were supported for AGdB and FC by the France 2030 program through the Idex Université Paris Cité (ANR-18-IDEX-0001).

\section*{In memorium}

This manuscript is dedicated to the memory of our colleagues and friends, Professor Serge Hazout to whom this special section is dedicated, and to Professor Narayanaswamy Srinivansan (1962-2021) who was an Indian molecular biophysicist and professor at the Molecular Biophysics Unit, Indian Institute of Science, Bangalore.

\section*{Funding}

This work has been supported by the Conseil Régional de La Réunion and Fonds Social Européen in the form of a PhD scholarship to SD under tier number 234275, convention number DIRED/20161451. BO is thankful to Conseil Régional Pays de la Loire for support in the framework of GRIOTE grant. PEACCEL was supported through a research program partially co-funded by the European Union (UE) and the Region Reunion (FEDER).
\vspace*{-12pt}

\section*{Competing interests}
FC is linked to Peaccel. 
SD, ST, RS, YHS, AGdB and BO declare no competing interests.

\bibliographystyle{elsarticle-num}
\bibliography{DHINGRA_BIOINFORMATICS_PAPER2/dhingra-bibliography}

\newpage

\section*{Supplementary Figures and Tables}

\subsection*{Supplementary Figures 1a-d. Coverage density plots}

The plots under this heading show the distribution of the number of fragments per position after Protein Block Prediction (PBP) for fragments with $atan\ score>=0.55$. The graphs are further classified into sections: (a) SCOP Class – all $\alpha$, (b) SCOP Class – all $\beta$, (c) SCOP Class - $\alpha$ + $\beta$ and (d) SCOP Class - $\alpha$ / $\beta$. The percentage of coverage for each test protein is marked above the plot. Along with it the positions for each protein with no fragment hits are marked in red and the number of residues with no hits is also shown above each graph.

\subsubsection*{Supplementary Figure 1a}
Protein Block Prediction (PBP) - SCOP Class – all $\alpha$\\
\includegraphics[width=0.95\textwidth]{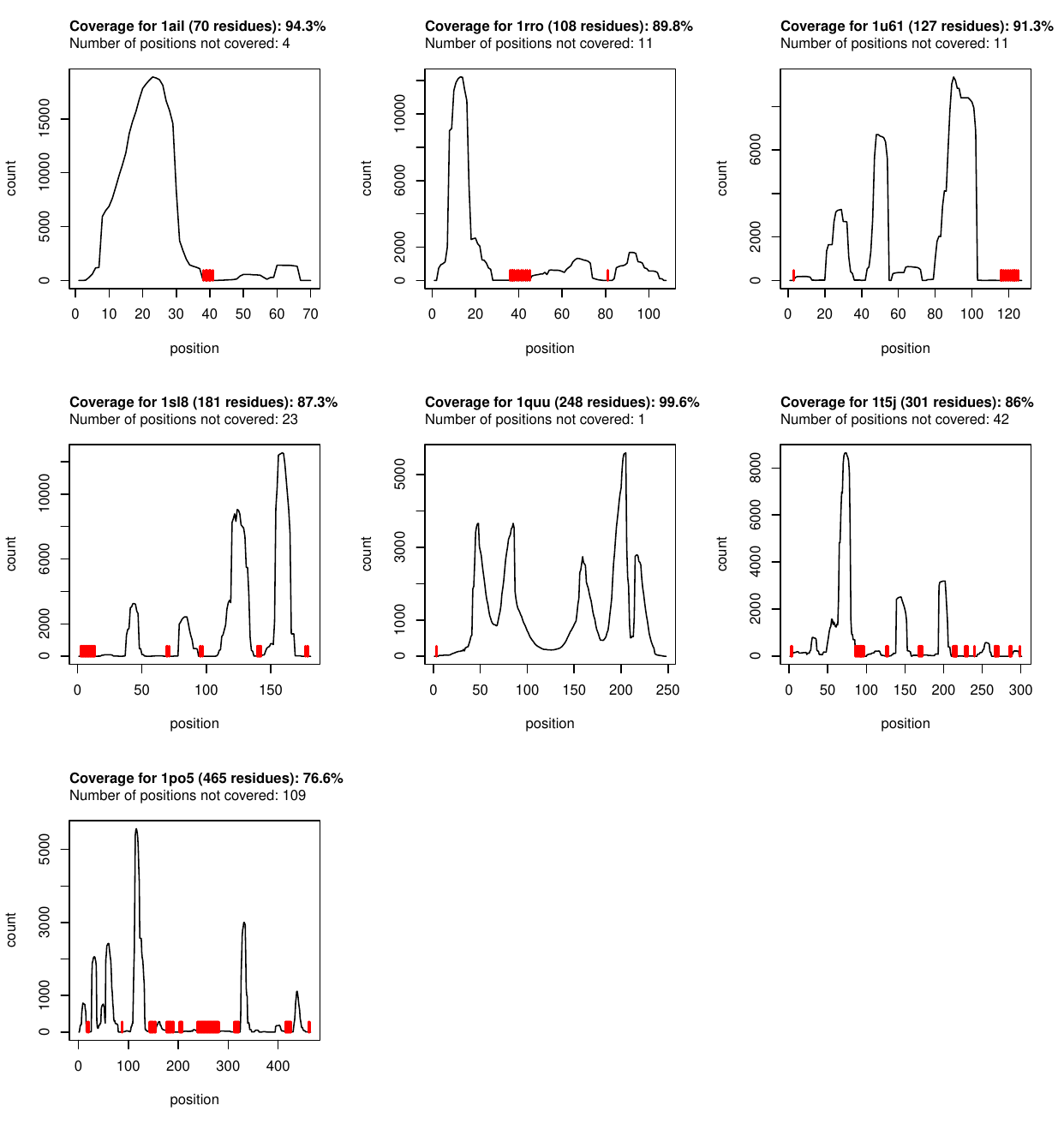}
\newpage
\subsubsection*{Supplementary Figure 1b}
Protein Block Prediction (PBP) - SCOP Class – all $\beta$\\
\includegraphics[width=0.95\textwidth]{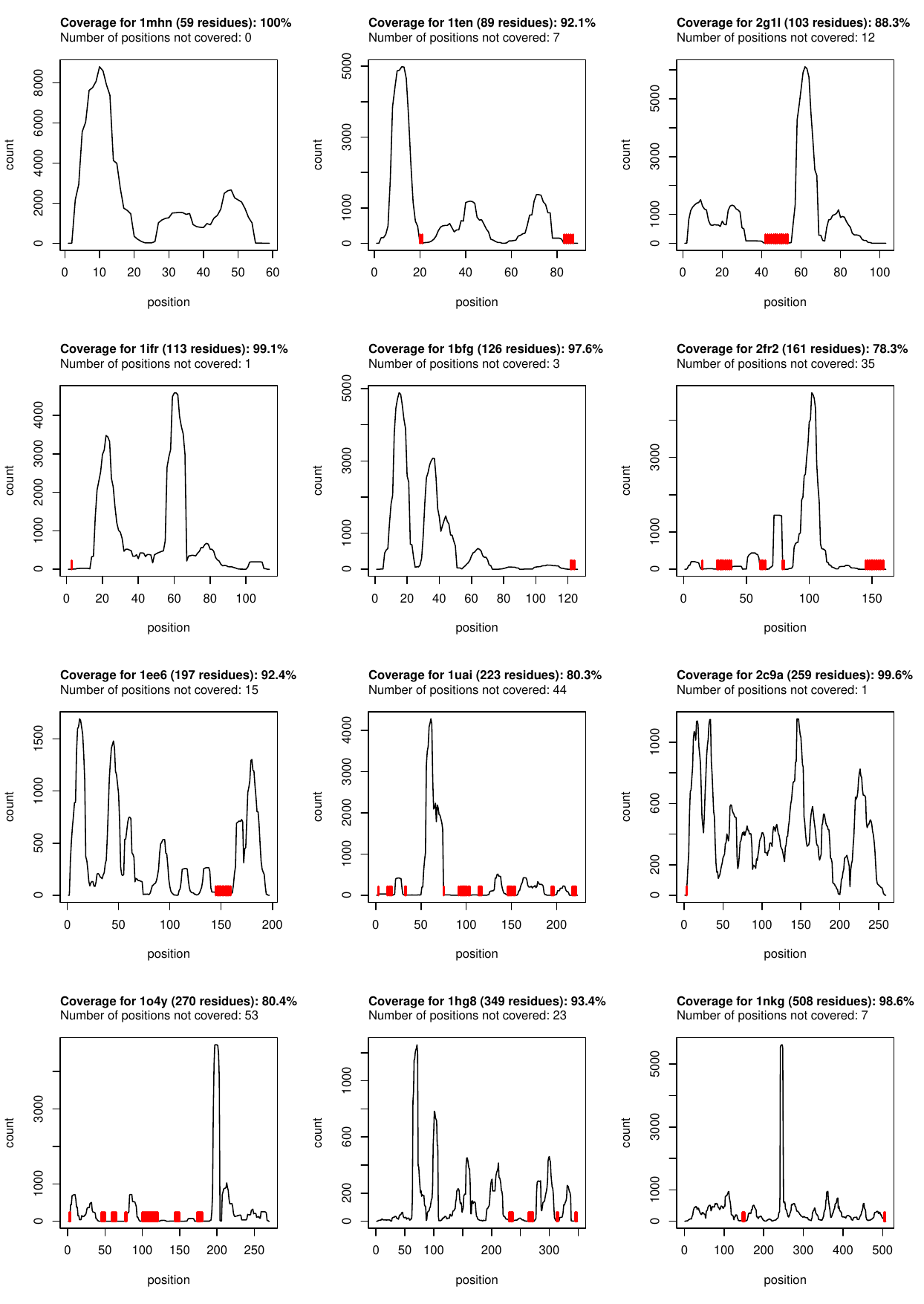}
\subsubsection*{Supplementary Figure 1c}
Protein Block Prediction (PBP) - SCOP Class – $\alpha$ + $\beta$\\
\includegraphics[width=0.95\textwidth]{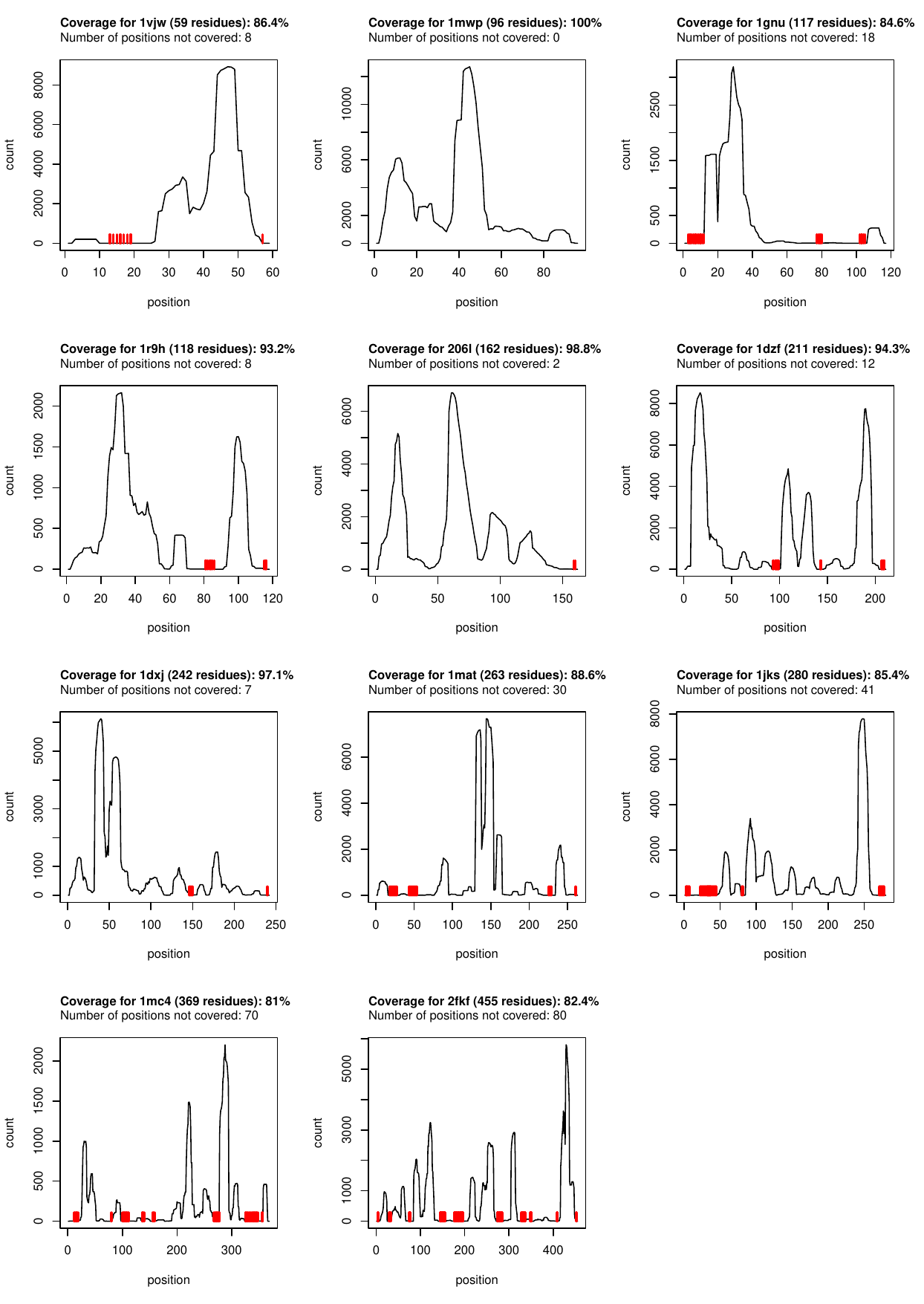}
\subsubsection*{Supplementary Figure 1d}
Protein Block Prediction (PBP) - SCOP Class – $\alpha$ / $\beta$\\
\includegraphics[width=0.95\textwidth]{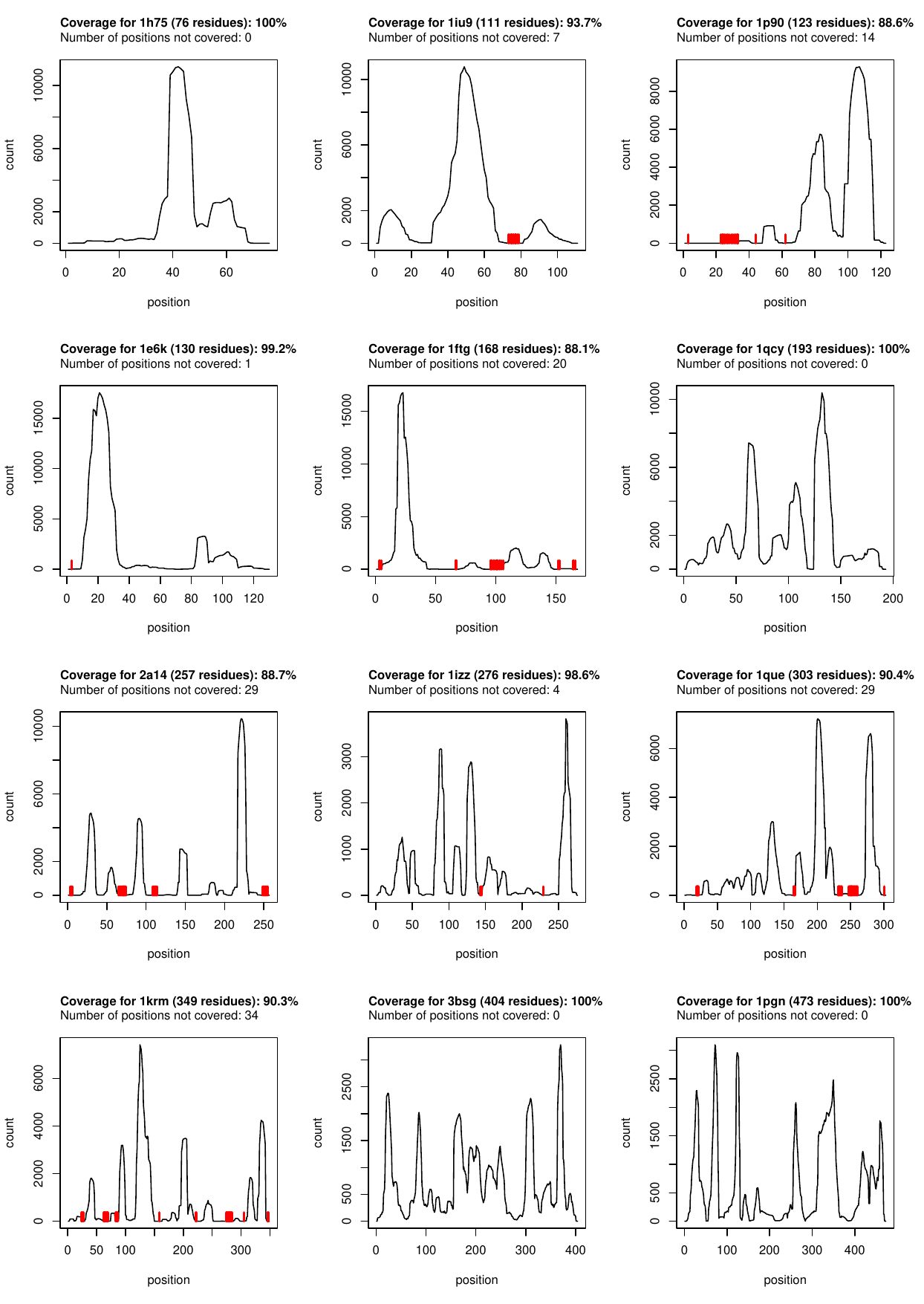}

\subsection*{Supplementary Figures 2. Sensitivity and specificity plots}

The plots under this heading show the co-relation between rmsd (2.5 \AA) and four chosen criteria for prioritizing fragment selection, i.e., protein sequence identity, protein sequence similarity, secondary structure identity and $atan\ score$. The analysis has been performed after Protein Block Prediction (PBP). The graphs are further classified into sections: (a) SCOP Class – all $\alpha$, (b) SCOP Class – all $\beta$, (c) SCOP Class - $\alpha$ + $\beta$ and (d) SCOP Class - $\alpha$ / $\beta$.

\subsubsection*{Supplementary Figure 2a. SCOP class: all $\alpha$}
\includegraphics[width=0.5\textwidth]{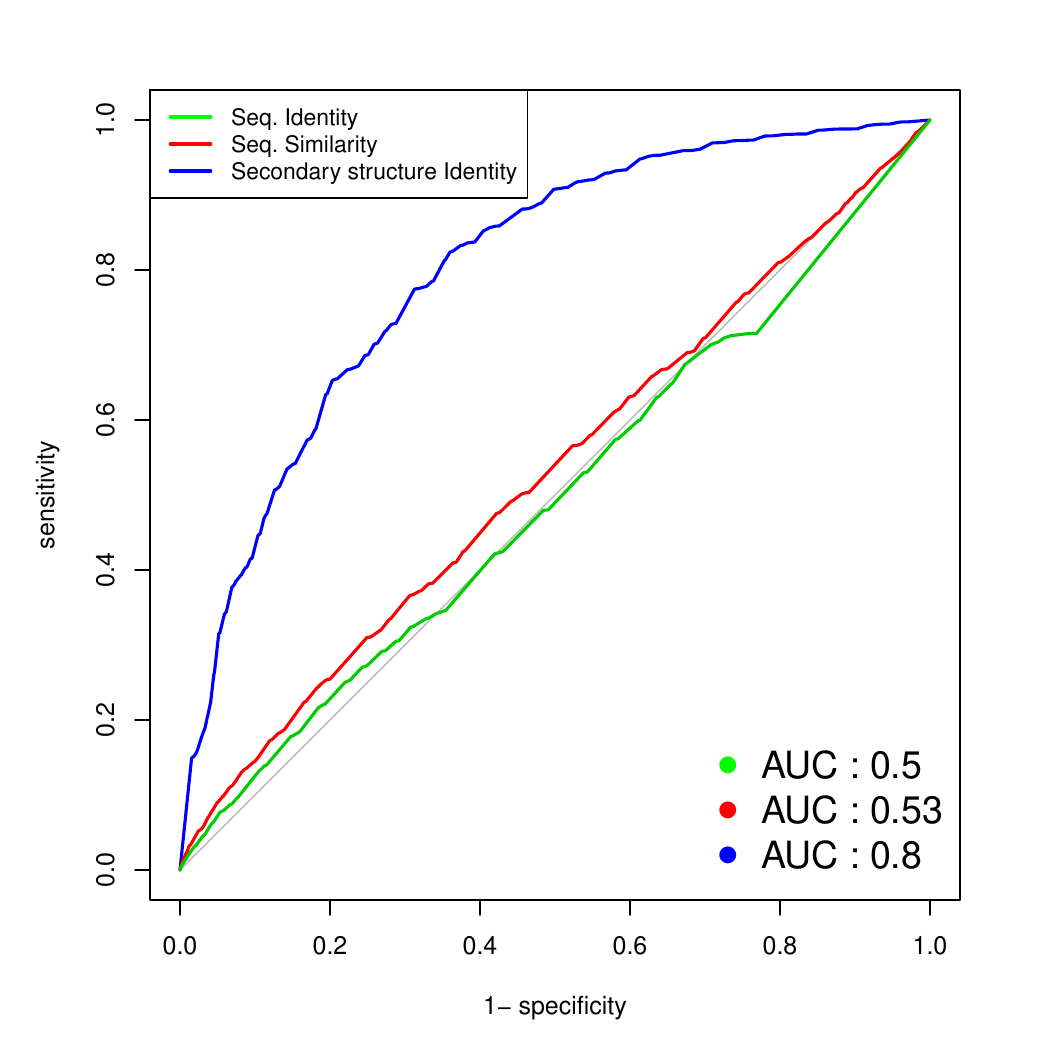}
\includegraphics[width=0.5\textwidth]{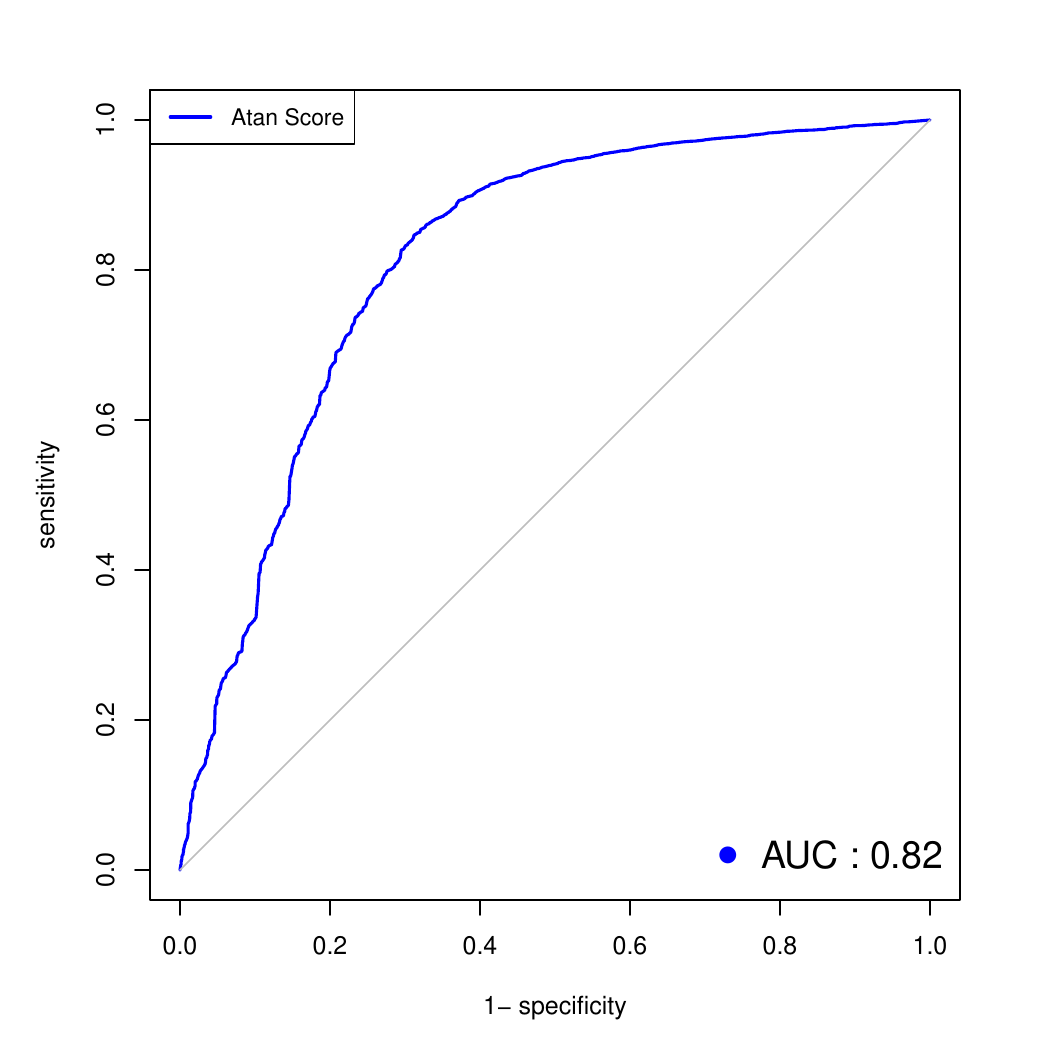}

\subsubsection*{Supplementary Figure 2b. SCOP class: all $\beta$}
\includegraphics[width=0.5\textwidth]{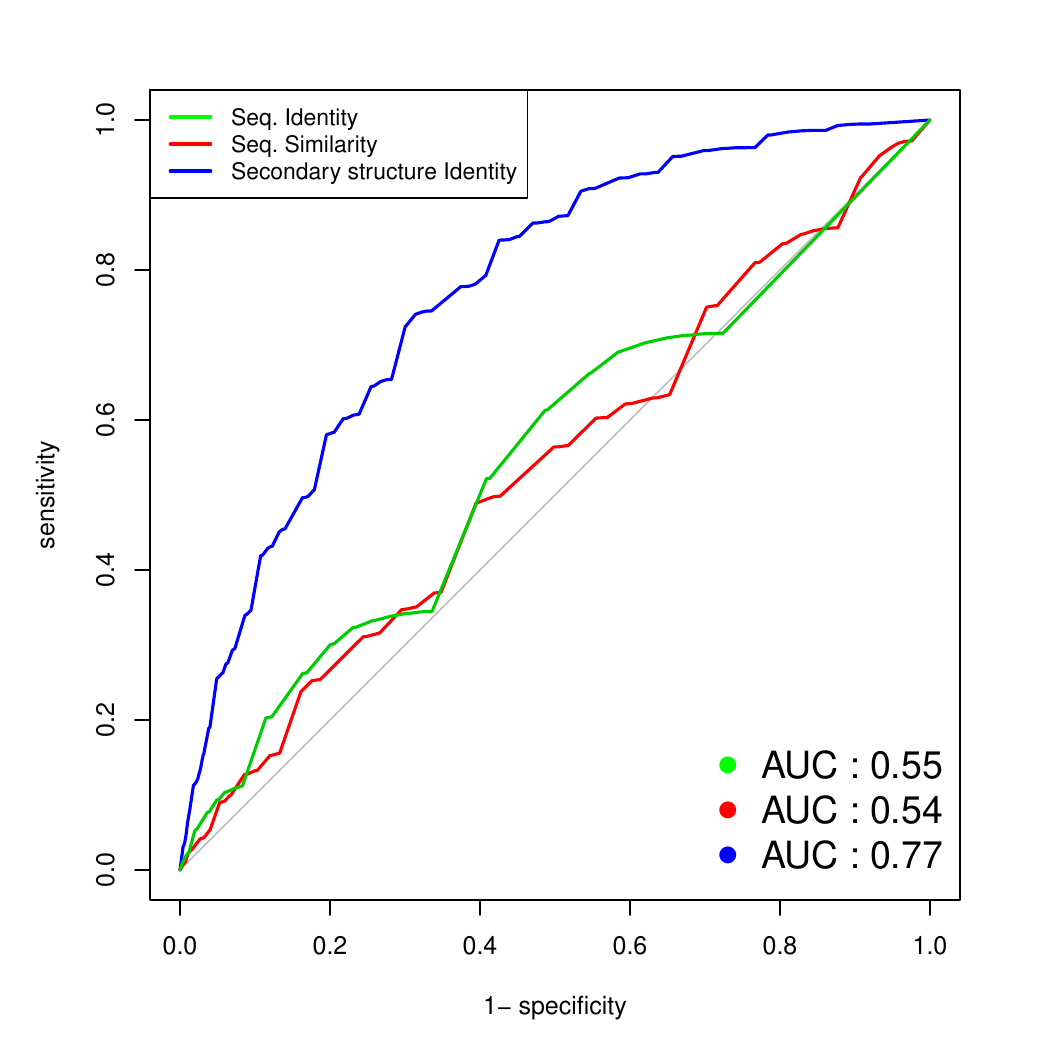}
\includegraphics[width=0.5\textwidth]{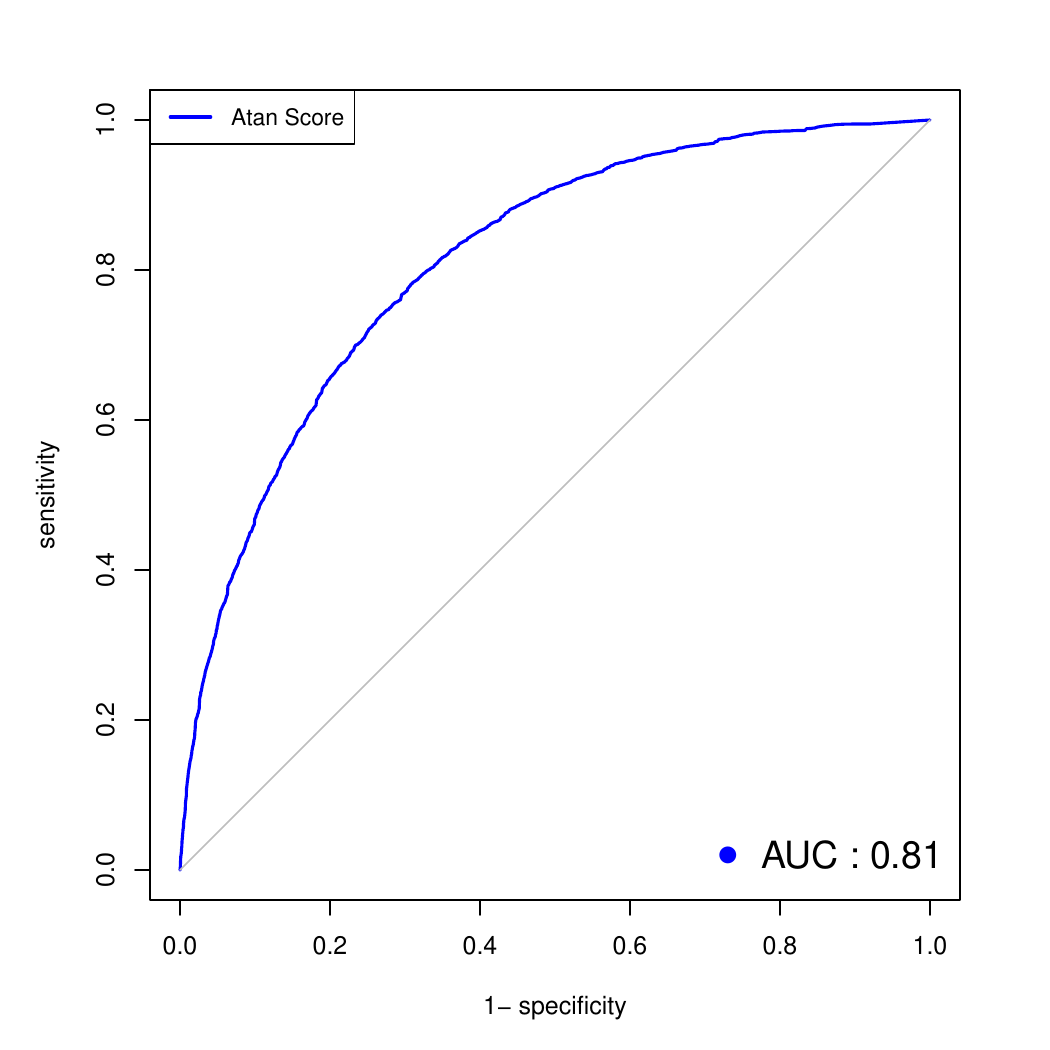}

\subsubsection*{Supplementary Figure 2c. SCOP class: $\alpha$ + $\beta$}
\includegraphics[width=0.5\textwidth]{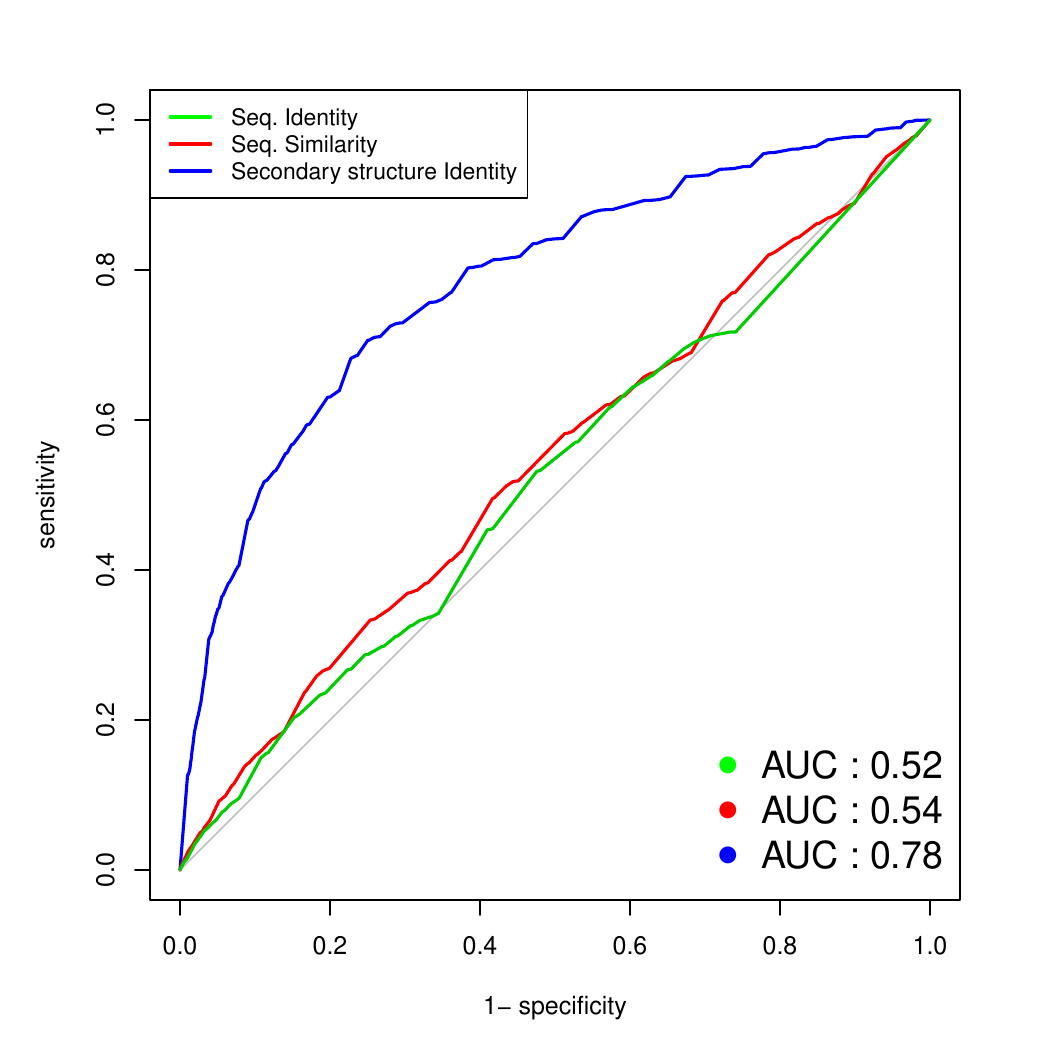}
\includegraphics[width=0.5\textwidth]{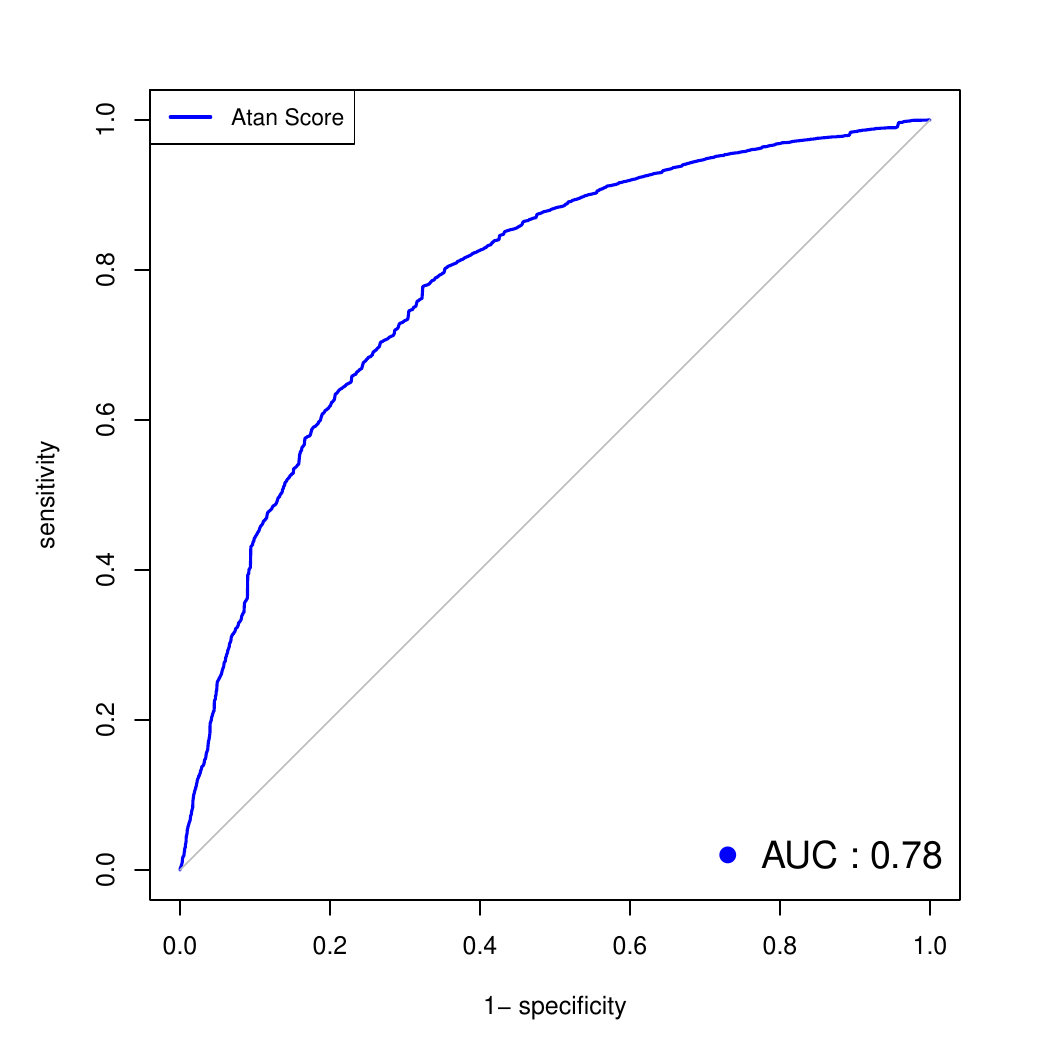}

\subsubsection*{Supplementary Figure 2d. SCOP class: $\alpha$ / $\beta$}
\includegraphics[width=0.5\textwidth]{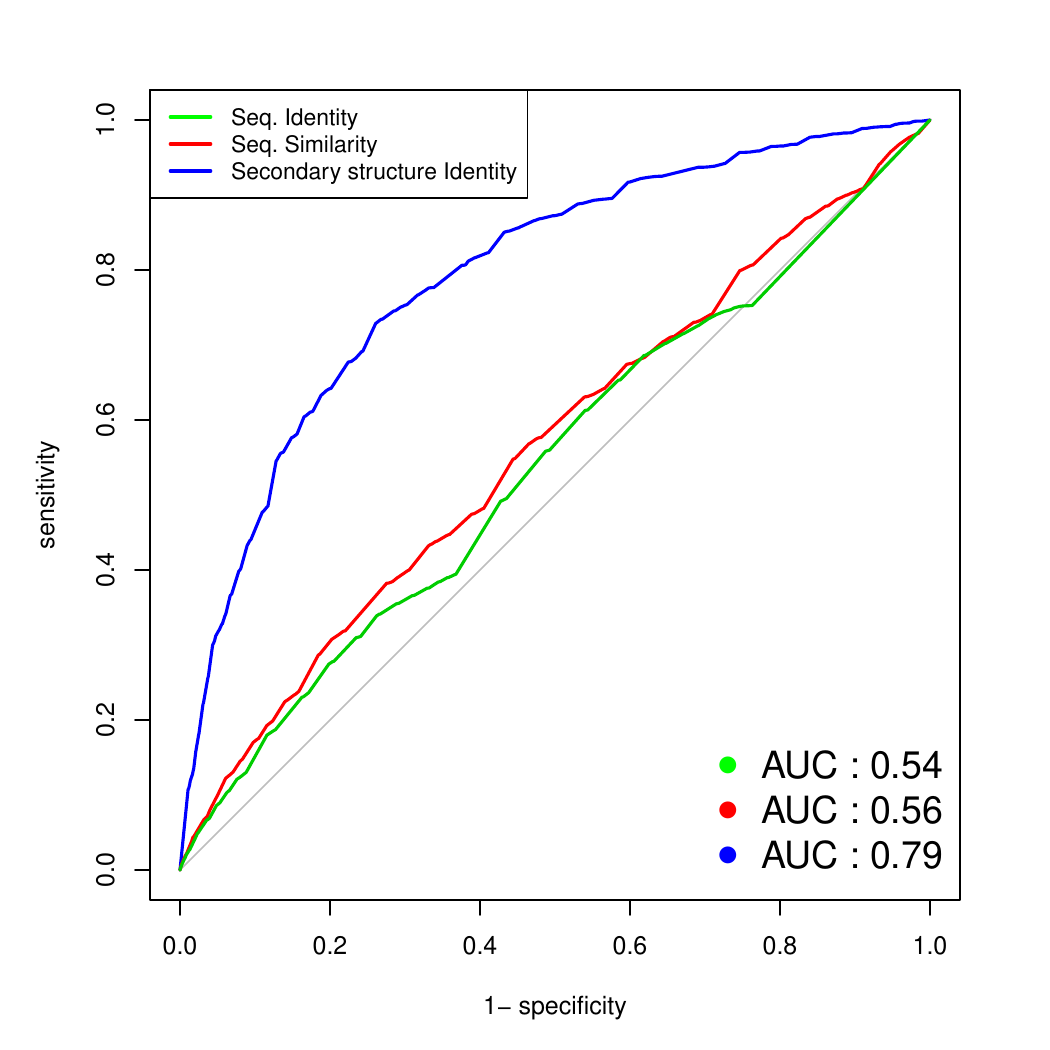}
\includegraphics[width=0.5\textwidth]{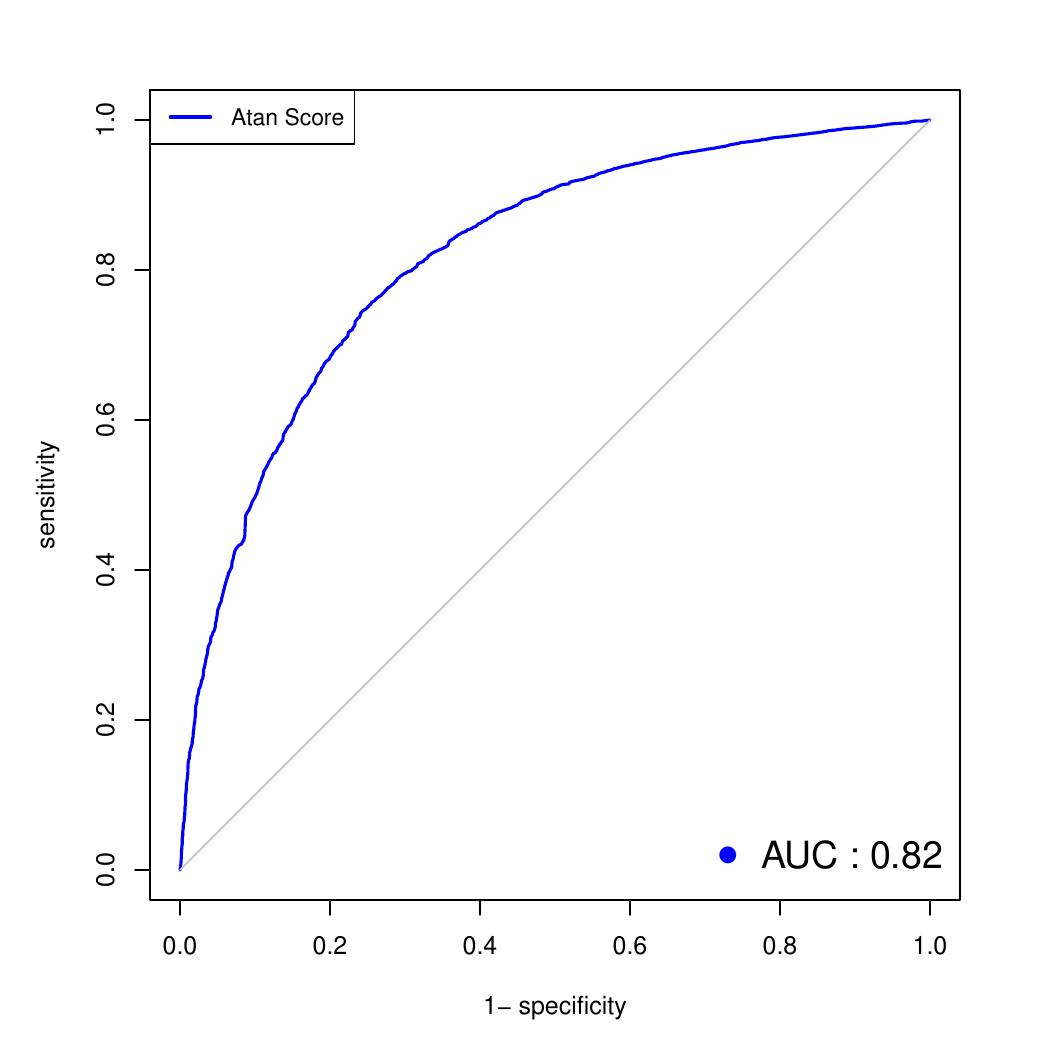}

\newpage
\subsection*{Supplementary Figures 3. \textit{RMSD} distribution per query protein}

Shown are the distributions of \textit{RMSD} values of fragments with $atan\ score\geq0.55$ starting at each position obtained with the pipeline. This illustrates the quality of the fragments.

\subsubsection*{Supplementary Figure 3a. Queries from all-$\alpha$ SCOP class.}
\includegraphics[page=1,width=1.0\textwidth]{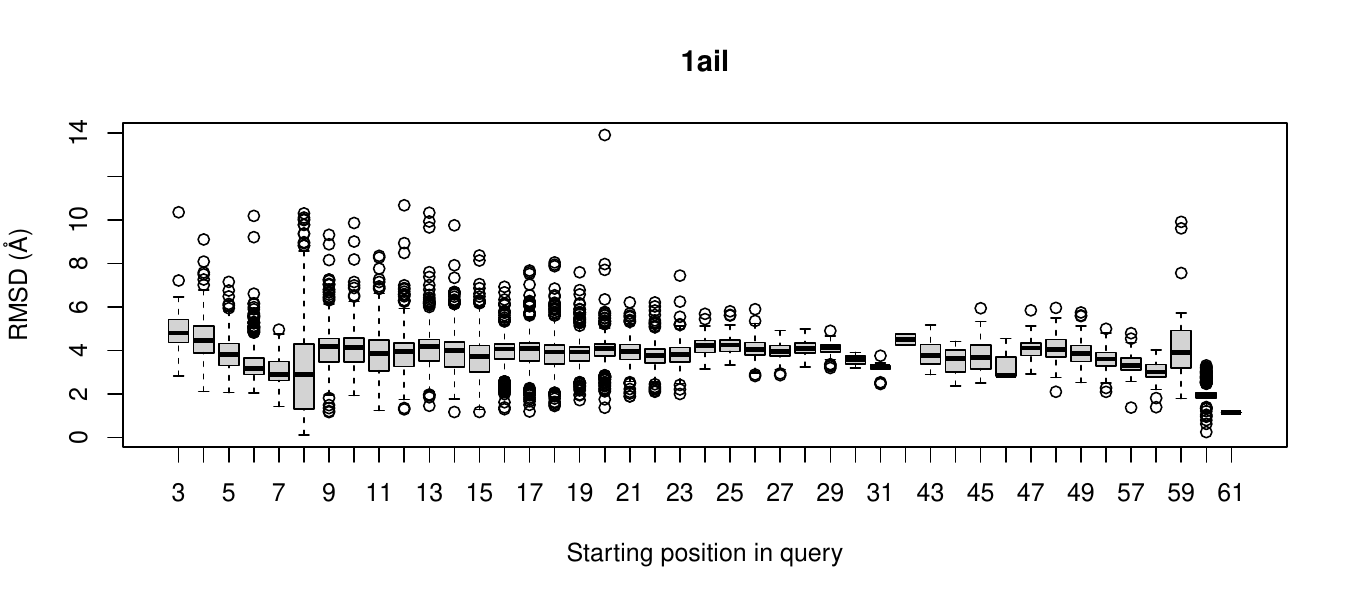}
\includegraphics[page=2,width=1.0\textwidth]{DHINGRA_BIOINFORMATICS_PAPER2/supplementary_materials/all_alpha_RMSD.pdf}
\includegraphics[page=3,width=1.0\textwidth]{DHINGRA_BIOINFORMATICS_PAPER2/supplementary_materials/all_alpha_RMSD.pdf}
\includegraphics[page=4,width=1.0\textwidth]{DHINGRA_BIOINFORMATICS_PAPER2/supplementary_materials/all_alpha_RMSD.pdf}
\includegraphics[page=5,width=1.0\textwidth]{DHINGRA_BIOINFORMATICS_PAPER2/supplementary_materials/all_alpha_RMSD.pdf}
\includegraphics[page=6,width=1.0\textwidth]{DHINGRA_BIOINFORMATICS_PAPER2/supplementary_materials/all_alpha_RMSD.pdf}
\includegraphics[page=7,width=1.0\textwidth]{DHINGRA_BIOINFORMATICS_PAPER2/supplementary_materials/all_alpha_RMSD.pdf}
\subsubsection*{Supplementary Figure 3b. Queries from all-$\beta$ SCOP class.}
\includegraphics[page=1,width=1.0\textwidth]{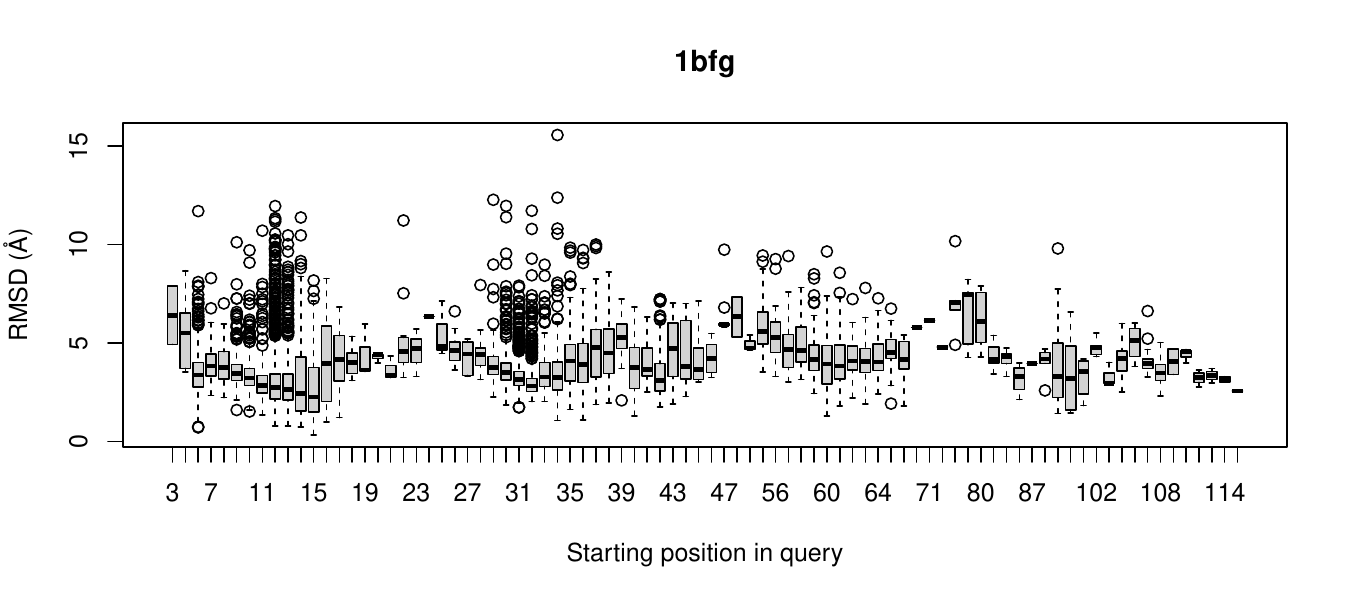}
\includegraphics[page=2,width=1.0\textwidth]{DHINGRA_BIOINFORMATICS_PAPER2/supplementary_materials/allbeta_RMSD.pdf}
\includegraphics[page=3,width=1.0\textwidth]{DHINGRA_BIOINFORMATICS_PAPER2/supplementary_materials/allbeta_RMSD.pdf}
\includegraphics[page=4,width=1.0\textwidth]{DHINGRA_BIOINFORMATICS_PAPER2/supplementary_materials/allbeta_RMSD.pdf}
\includegraphics[page=5,width=1.0\textwidth]{DHINGRA_BIOINFORMATICS_PAPER2/supplementary_materials/allbeta_RMSD.pdf}
\includegraphics[page=6,width=1.0\textwidth]{DHINGRA_BIOINFORMATICS_PAPER2/supplementary_materials/allbeta_RMSD.pdf}
\includegraphics[page=7,width=1.0\textwidth]{DHINGRA_BIOINFORMATICS_PAPER2/supplementary_materials/allbeta_RMSD.pdf}
\includegraphics[page=8,width=1.0\textwidth]{DHINGRA_BIOINFORMATICS_PAPER2/supplementary_materials/allbeta_RMSD.pdf}
\includegraphics[page=9,width=1.0\textwidth]{DHINGRA_BIOINFORMATICS_PAPER2/supplementary_materials/allbeta_RMSD.pdf}
\includegraphics[page=10,width=1.0\textwidth]{DHINGRA_BIOINFORMATICS_PAPER2/supplementary_materials/allbeta_RMSD.pdf}
\includegraphics[page=11,width=1.0\textwidth]{DHINGRA_BIOINFORMATICS_PAPER2/supplementary_materials/allbeta_RMSD.pdf}
\includegraphics[page=12,width=1.0\textwidth]{DHINGRA_BIOINFORMATICS_PAPER2/supplementary_materials/allbeta_RMSD.pdf}
\subsubsection*{Supplementary Figure 3c. Queries from $\alpha + \beta$ SCOP class.}
\includegraphics[page=1,width=1.0\textwidth]{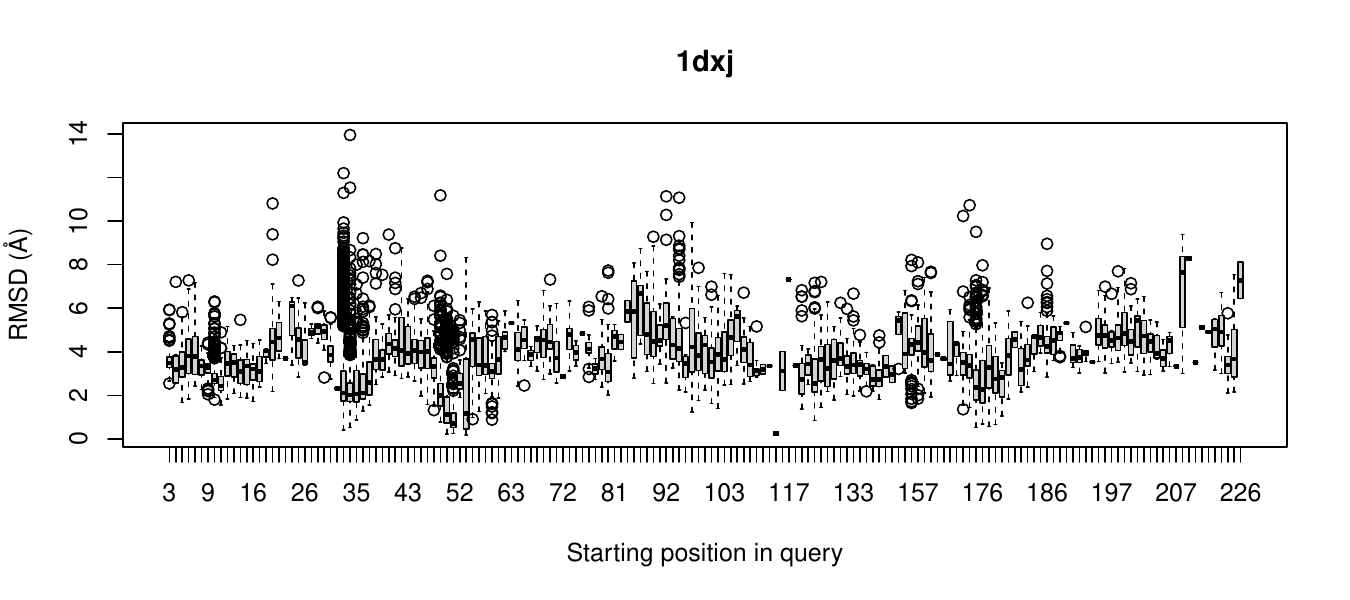}
\includegraphics[page=2,width=1.0\textwidth]{DHINGRA_BIOINFORMATICS_PAPER2/supplementary_materials/alphaAbeta_RMSD.pdf}
\includegraphics[page=3,width=1.0\textwidth]{DHINGRA_BIOINFORMATICS_PAPER2/supplementary_materials/alphaAbeta_RMSD.pdf}
\includegraphics[page=4,width=1.0\textwidth]{DHINGRA_BIOINFORMATICS_PAPER2/supplementary_materials/alphaAbeta_RMSD.pdf}
\includegraphics[page=5,width=1.0\textwidth]{DHINGRA_BIOINFORMATICS_PAPER2/supplementary_materials/alphaAbeta_RMSD.pdf}
\includegraphics[page=6,width=1.0\textwidth]{DHINGRA_BIOINFORMATICS_PAPER2/supplementary_materials/alphaAbeta_RMSD.pdf}
\includegraphics[page=7,width=1.0\textwidth]{DHINGRA_BIOINFORMATICS_PAPER2/supplementary_materials/alphaAbeta_RMSD.pdf}
\includegraphics[page=8,width=1.0\textwidth]{DHINGRA_BIOINFORMATICS_PAPER2/supplementary_materials/alphaAbeta_RMSD.pdf}
\includegraphics[page=9,width=1.0\textwidth]{DHINGRA_BIOINFORMATICS_PAPER2/supplementary_materials/alphaAbeta_RMSD.pdf}
\includegraphics[page=10,width=1.0\textwidth]{DHINGRA_BIOINFORMATICS_PAPER2/supplementary_materials/alphaAbeta_RMSD.pdf}
\includegraphics[page=11,width=1.0\textwidth]{DHINGRA_BIOINFORMATICS_PAPER2/supplementary_materials/alphaAbeta_RMSD.pdf}
\subsubsection*{Supplementary Figure 3d. Queries from $\alpha / \beta$ SCOP class.}
\includegraphics[page=1,width=1.0\textwidth]{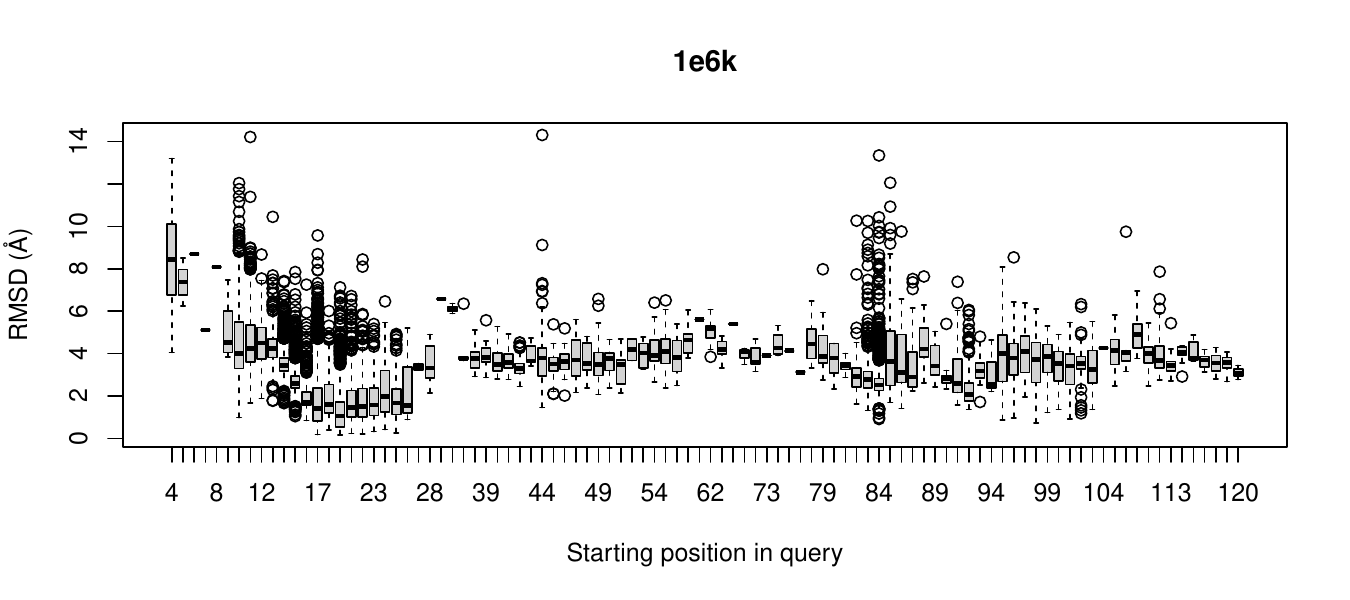}
\includegraphics[page=2,width=1.0\textwidth]{DHINGRA_BIOINFORMATICS_PAPER2/supplementary_materials/alphaObeta_RMSD.pdf}
\includegraphics[page=3,width=1.0\textwidth]{DHINGRA_BIOINFORMATICS_PAPER2/supplementary_materials/alphaObeta_RMSD.pdf}
\includegraphics[page=4,width=1.0\textwidth]{DHINGRA_BIOINFORMATICS_PAPER2/supplementary_materials/alphaObeta_RMSD.pdf}
\includegraphics[page=5,width=1.0\textwidth]{DHINGRA_BIOINFORMATICS_PAPER2/supplementary_materials/alphaObeta_RMSD.pdf}
\includegraphics[page=6,width=1.0\textwidth]{DHINGRA_BIOINFORMATICS_PAPER2/supplementary_materials/alphaObeta_RMSD.pdf}
\includegraphics[page=7,width=1.0\textwidth]{DHINGRA_BIOINFORMATICS_PAPER2/supplementary_materials/alphaObeta_RMSD.pdf}
\includegraphics[page=8,width=1.0\textwidth]{DHINGRA_BIOINFORMATICS_PAPER2/supplementary_materials/alphaObeta_RMSD.pdf}
\includegraphics[page=9,width=1.0\textwidth]{DHINGRA_BIOINFORMATICS_PAPER2/supplementary_materials/alphaObeta_RMSD.pdf}
\includegraphics[page=10,width=1.0\textwidth]{DHINGRA_BIOINFORMATICS_PAPER2/supplementary_materials/alphaObeta_RMSD.pdf}
\includegraphics[page=11,width=1.0\textwidth]{DHINGRA_BIOINFORMATICS_PAPER2/supplementary_materials/alphaObeta_RMSD.pdf}
\includegraphics[page=12,width=1.0\textwidth]{DHINGRA_BIOINFORMATICS_PAPER2/supplementary_materials/alphaObeta_RMSD.pdf}
\newpage
\subsection*{Supplementary Tables}
\subsubsection*{Supplementary Table 1.}
 This table provides the fragment hit counts and coverage obtained after Protein Block Prediction (PBP) for each protein from the query dataset.

 \includegraphics[width=0.95\textwidth]{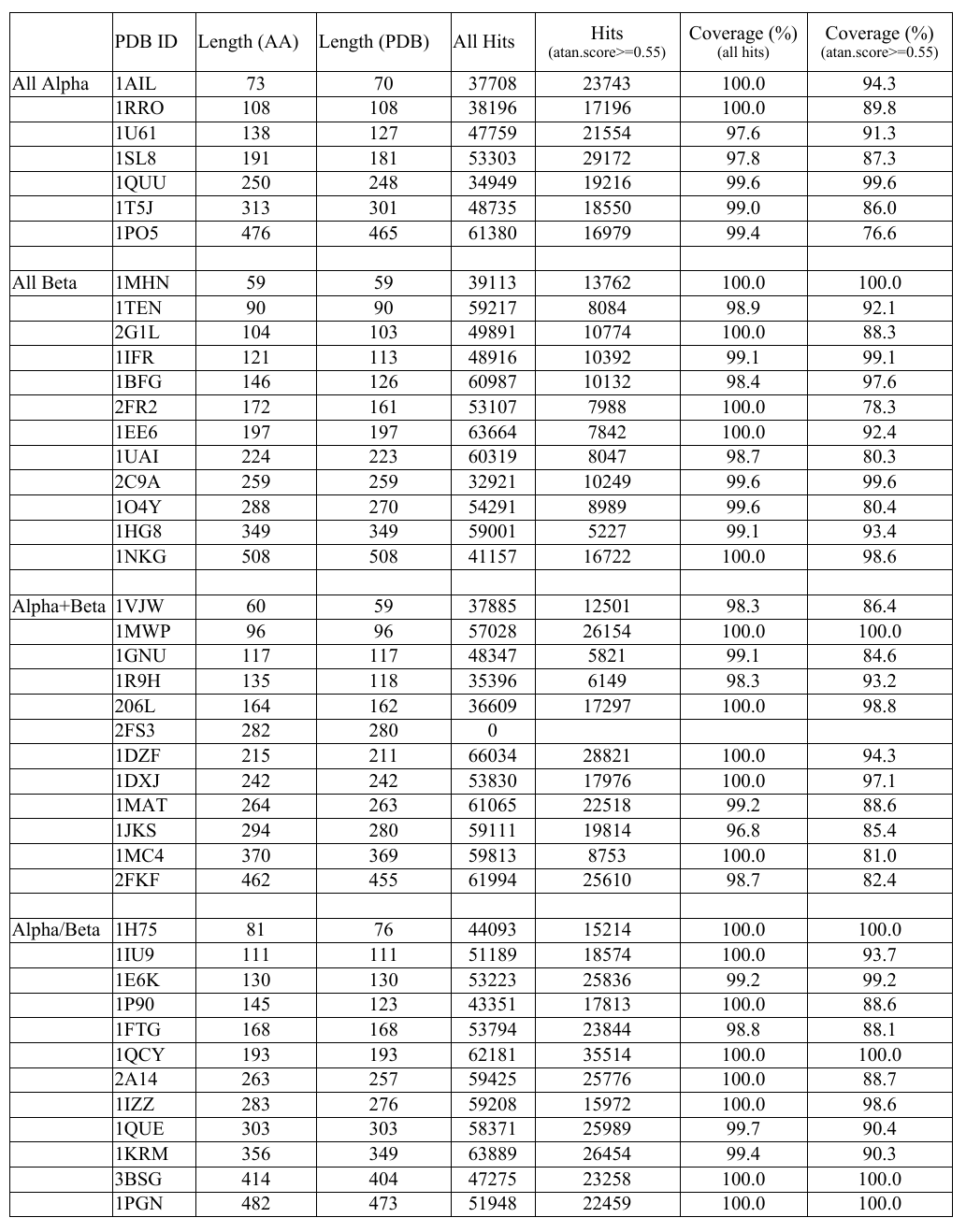}
 \newpage
 \subsubsection*{Supplementary Table 2.}
Statistics of the length of all the fragments obtained after Protein Block Prediction (PBP) for each protein from the query dataset.

 \includegraphics[width=0.95\textwidth]{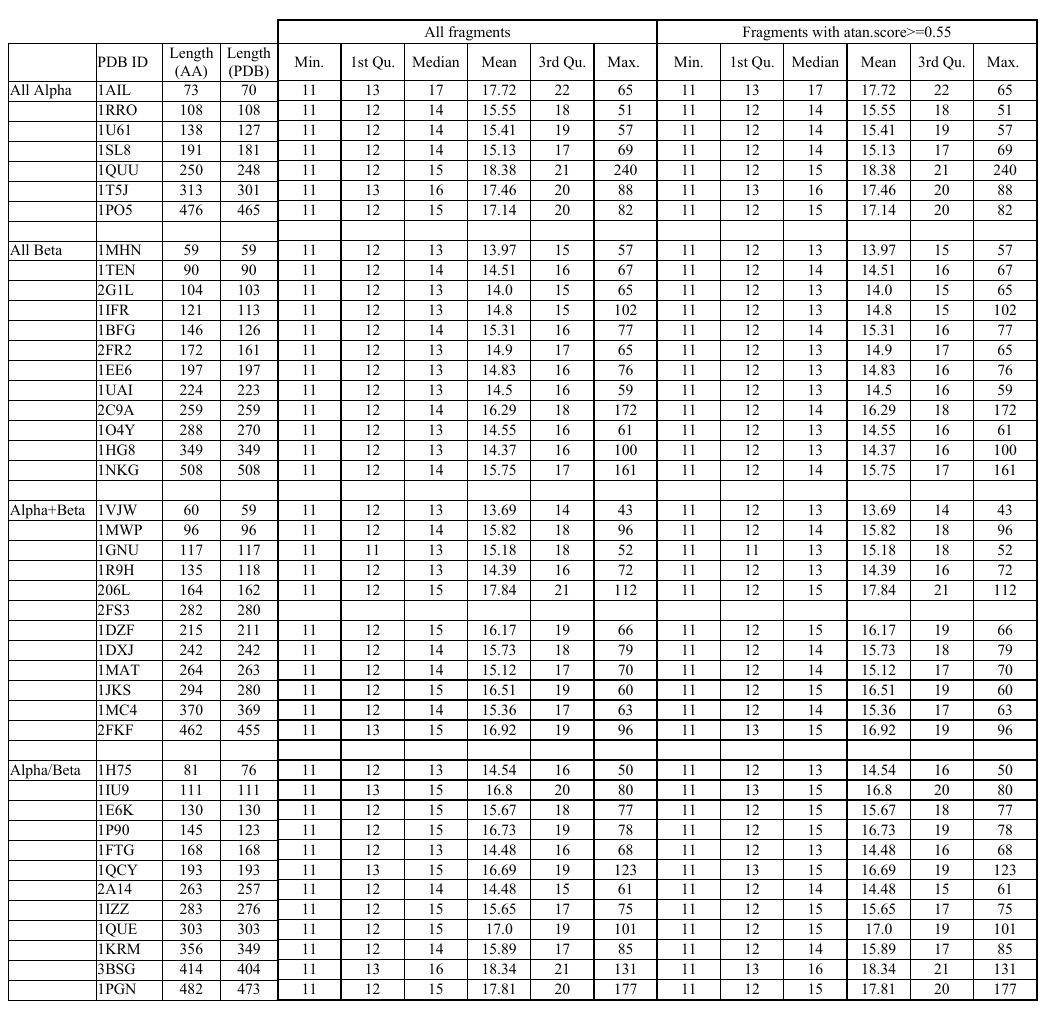}

 \newpage
\subsubsection*{Supplementary Table 3.}
Quantification of precision for regions within and outside of regular secondary structures after protein blocks predicted (PBP) and filtering of fragments with $atan\ score >=0.55$. Shown are mean and standard deviation for \textit{RMSD} values per type of local regular secondary structure.

\includegraphics[width=0.85\textwidth]{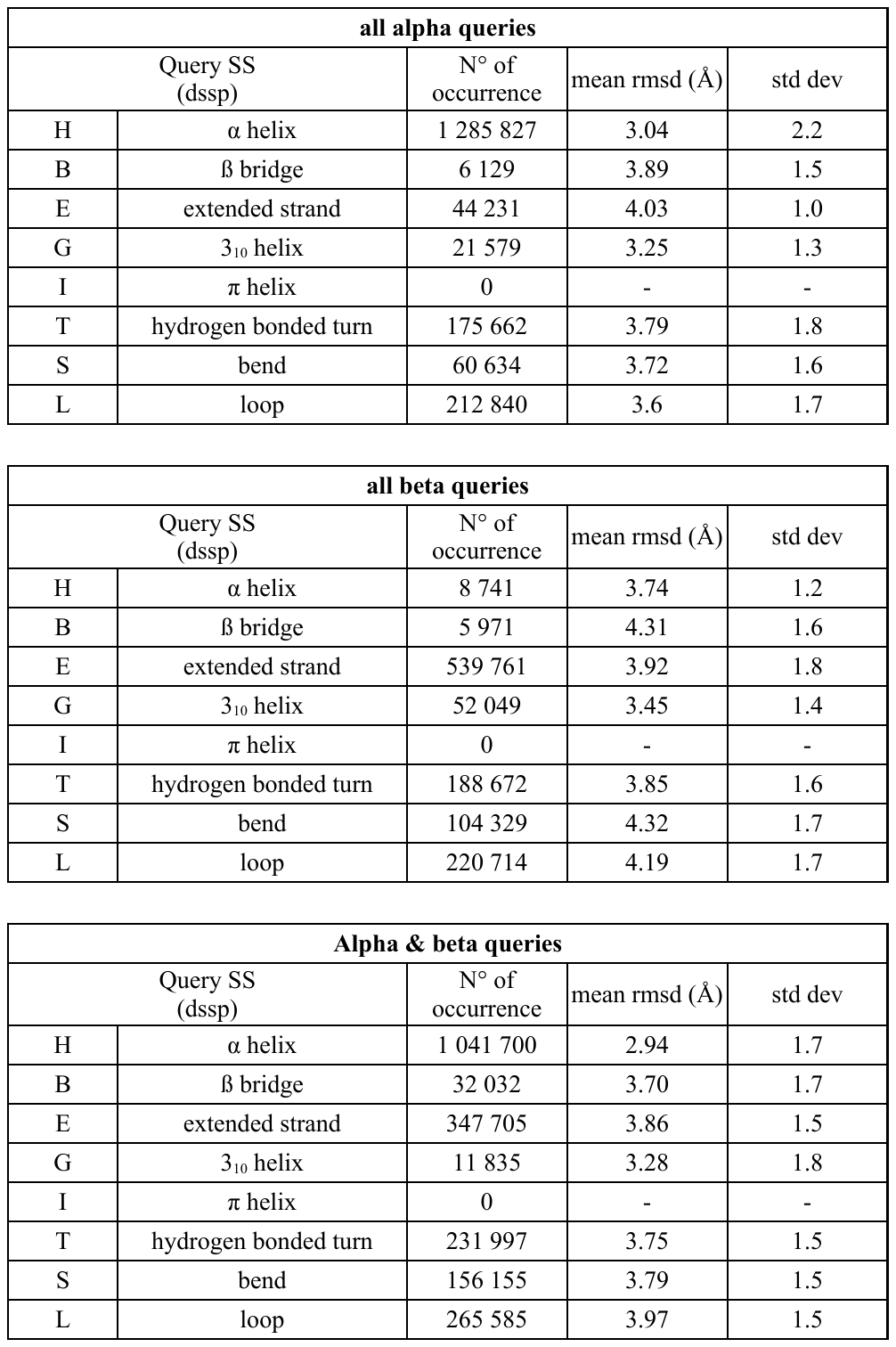}

\subsubsection*{Supplementary Table 3 (continued).}
\includegraphics[width=0.85\textwidth]{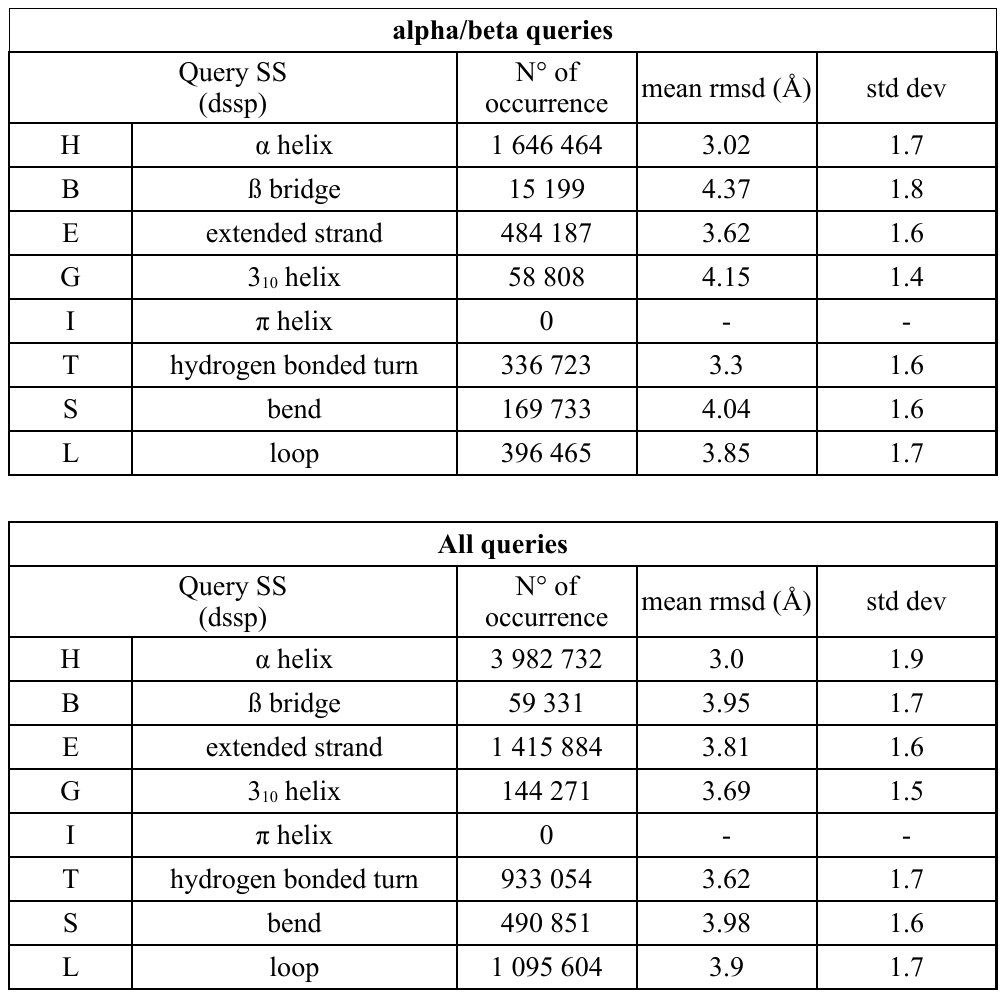}




\end{document}